\documentclass[12pt]{article}

\usepackage[colorlinks,linkcolor=.]{hyperref}
\usepackage{cite}
\usepackage{amsmath}
\usepackage{bm}
\usepackage{amssymb}
\usepackage{amsthm}
\usepackage[table]{xcolor}

\usepackage{bbm}
\usepackage{slashed}

\newcommand{\avg}[1]{\left\langle#1\right\rangle}

\newcommand{\parens}[1]{\!\left(#1\right)}

\newcommand{\brackets}[1]{\!\left[#1\right]}

\def\be{\begin{equation}}
\def\ee{\end{equation}}

\usepackage{amsfonts}
\usepackage{amssymb}
\usepackage{mathtools}
\usepackage{graphicx}
\usepackage{wrapfig}
\usepackage{caption}
\usepackage{feynmp-auto}

\DeclareMathOperator{\Spin}{Spin}

\DeclareMathOperator{\tra}{tr}
\DeclareMathOperator{\Real}{Re}

\newcommand{\Li}{\mathcal{L}}
\newcommand{\M}{\mathcal{M}}

\newcommand{\Oo}{\mathcal{O}}

\makeatletter
\newcommand*\wt[1]{\mathpalette\wthelper{#1}}
\newcommand*\wthelper[2]{%
        \hbox{\dimen@\accentfontxheight#1%
                \accentfontxheight#11.15\dimen@
                $\m@th#1\widetilde{#2}$%
                \accentfontxheight#1\dimen@
        }%
}

\newcommand*\accentfontxheight[1]{%
        \fontdimen5\ifx#1\displaystyle
                \textfont
        \else\ifx#1\textstyle
                \textfont
        \else\ifx#1\scriptstyle
                \scriptfont
        \else
                \scriptscriptfont
        \fi\fi\fi3
}
\makeatother

\newcommand*{\sgn}{\ensuremath{\mathrm{sgn}}}

\newcommand*{\momentumarrow}[4]{%
    \fmfcmd{style_def marrow#1
    expr p = drawarrow subpath (0.3, 0.7) of p shifted #3 #2 withpen pencircle scaled 0.4;
    enddef;}
    \fmf{marrow#1,tension=0}{#4}}

\begin{document}
\unitlength = 1mm

\title{\bf Strong-Weak Chern-Simons-Matter Dualities \\ from a Lattice Construction}
\author{Jing-Yuan Chen and Max Zimet \\ {\small\em Stanford Institute for Theoretical Physics, Stanford University, Stanford, CA 94305, USA}}

\date{}
\maketitle

\begin{abstract}
We provide a lattice demonstration of $(2+1)$-dimensional field theory dualities relating free Dirac or Majorana fermions to strongly-interacting bosonic Chern-Simons-matter theories. Specifically, we prove the recent conjecture that $U(N)$ level-1 with $N_f$ gauged complex Wilson-Fisher scalars (where $1\le N_f\le N$) is dual to $N_f$ Dirac fermions, as well as the analogous conjecture relating $SO(N)$ theories with real Wilson-Fisher scalars to Majorana fermions for $1\le N_f\le N-2$. Furthermore, we discover new dualities that allow us to explain the interesting phase structure of the $SO(N)$ theories with $N-1$ and $N$ scalars, for all $N\ge 2$.
\end{abstract}

\tableofcontents

\section{Introduction and Conclusion}

Recently, new (2+1)-dimensional field theory dualities -- including boson-fermion dualities (generalizing flux attachment in the condensed matter literature) -- have been under intensive study.\footnote{Relativistic versions of flux attachment are, in fact, an old idea \cite{Polyakov:1988md}. The novelty in recent proposals is that this extends even to conformal fixed points.} They have roots in large $N$ studies of models dual to Vasiliev gravity \cite{AharonyVasiliev, YinVasiliev, AharonyVasiliev2}, as well as attempts to understand the physics of the fractional quantum Hall system and topological insulators \cite{Son:2015xqa, Wang:2015qmt, Metlitski:2015eka}. Many of these dualities were conjectured in \cite{Aharony:2015mjs}. The simplest boson/fermion dualities were then crystallized in \cite{Karch:2016sxi, Seiberg:2016gmd}, while dualities with non-Abelian gauge groups were further studied in \cite{Hsin:2016blu} and \cite{Metlitski:2016dht, Aharony:2016jvv}, which respectively focused on unitary and orthogonal gauge groups. These simple dualities serve to generate a larger web of dualities, for instance by using the natural $SL(2,Z)$ action on (2+1)-dimensional conformal field theories with Abelian global currents \cite{Witten}. Additional recent conjectures and tests of dualities include \cite{Xu:2015lxa, Benini:2017dus, Giombi:2017txg, Gaiotto:2017tne, Gomis:2017ixy, Cordova:2017vab, Cordova:2017kue, Karch:2016aux, Jensen:2017dso, Benini:2017aed, Jensen:2017bjo, Aitken:2017nfd,Aitken:2018joi}, and recent condensed matter applications of these dualities include \cite{Murugan:2016zal, Wang:2017txt, Qin:2017cqw, Janssen:2017eeu, You:2017mkc, Goldman:2017zep, Hui:2017pwe, Hui:2017cyz}.

Heuristic derivations have appeared using wire constructions \cite{Mross:2015idy, Mross:2017gny}, deformations of well-established supersymmetric dualities \cite{minwalla:SeibergDual, Kachru:2015rma, Yacobi:SeibergDual, Kachru:2016rui, SUSY2}, holography \cite{Jensen:2017xbs}, loop models \cite{goldman:loops}, and an exact lattice duality \cite{Chen:2017lkr}. In this paper, we generalize the lattice construction of \cite{Chen:2017lkr} in order to study non-Abelian theories with multiple matter flavors. (As in all of the `derivations' we have mentioned, we will need to make some weak assumptions about what our theories flow to in the infrared. However, our assumptions are very weak, and in many cases, including the most interesting cases where we provide new dualities, they amount only to the assumption that our lattice theories flow to their obvious continuum counterparts.\footnote{See Appendix \ref{sec:phaseTransitions} for a more detailed discussion of the extent to which our assumptions are innocuous.}) We have a number of motivations for doing so. First, while the dualities of interest formally arise from the more general conjecture \cite{Aharony:2015mjs}
\be SU(K)_{-N+\frac{N_f}{2}} + N_f \,\, \mbox{Dirac fermions} \quad \longleftrightarrow \quad U(N)_K + N_f \,\, \mbox{complex Wilson-Fisher scalars} \label{eq:Udual} \ee
and its $SO/SO$ counterpart (with Majorana fermions and real scalars) by setting $K=1$, since $SU(1)$ and $SO(1)$ are trivial, they are nevertheless rather surprising, as one side is independent of $N$ while the other is not. This aspect of the dualities played an important role in the recent applications of \cite{Hui:2017pwe, Hui:2017cyz}. Second, the lattice non-linear sigma model proves to be an elegant description of the Wilson-Fisher theories, as it accounts for all of the universal behavior in the potentials of \cite{Hsin:2016blu,Aharony:2016jvv} while eliminating the irrelevant radial modes of the scalars. Additionally, as we explain below, the lattice is a powerful tool for obtaining dualities, and it is important to see how far this technique can be developed. In fact, we will provide interesting new dualities\footnote{As we discuss in Appendix \ref{sec:phaseTransitions}, strictly speaking we are able to prove new dualities for fixed points involving scalars coupled to Chern-Simons, but calling them `gauged Wilson-Fisher' fixed points might be presumptuous.} in the $SO/SO$ case when $0 \le N-N_f \le 1$. The phase structure of the latter theories is non-trivial (see figures \ref{fig:nm1} and \ref{fig:n} in section \ref{sec:Nf}) and depends on $N$ in interesting ways that are difficult to discern without explicit calculations such as those that appear below. In the future, we hope to be able to provide evidence for and nail down some of the details in the proposals of \cite{Komargodski:2017keh}.

While we defer a detailed description of our lattice proof to the body of the paper, we wish to emphasize here the main reasons why the lattice construction is powerful. To an IR field theorist, Chern-Simons-matter theories are intractable strongly-coupled systems (excepting certain limiting values of the parameters). However, we can obtain a Chern-Simons interaction by beginning in the UV with a massive fermion. (Indeed, this approach allows us to guarantee that we obtain the correct dependence on the gravitational and electromagnetic backgrounds, as well as the topology, in the IR.) The idea is then to integrate out the gauge field and scalars and demonstrate that the resulting theory describes free fermions in the infrared. Integrating out the bosons will generate interactions for the fermions, so one might fear that one loses control in the infrared. In fact, one might suspect that the resulting theory would be highly non-local, since we are integrating out massless bosons. However, we only have critical bosons in the IR; in the UV, the Higgs mechanism and confinement together prevent us from ever having to integrate out light bosons, and so we are able to find a local fermionic theory. That this is possible is ultimately due to the existence of the duality. Confinement results from our setting the Maxwell coupling to infinity \emph{at the lattice scale}. One might wonder about the IR description of a gauge field which has no kinetic term in the UV, but the parity anomaly and the paucity of relevant operators strongly suggest that a level one Chern-Simons-matter theory obtains in the IR.\footnote{In fact, although we set the Maxwell coupling to infinity, our derivation makes clear that -- thanks to the Higgs mechanism -- for the most part only small fluctuations of the gauge field play a role, so the important gauge field path integrals are performed only over the Lie algebra, and there is no question that our theories are the appropriate lattice avatars of the continuum theories of interest. The exception to this rule is that in some cases we will need to assume that certain theories with large gauge field fluctuations confine with a mass gap, and when we do so we assume that the analogous statement also holds for the continuum theory.

Indeed, one could easily retain the Maxwell interaction with a large coupling constant, $e^2$, but it would not change anything, as its effects would be suppressed by $T/e^2$, the inverse of the square of the Higgs scale, as is evident from the modified propagator. We demonstrate in Appendix \ref{sec:YMcomputation} that a small $e^2$ is also tractable.\label{ft:gaugeField}}

Of course, there are other coupling constants in the IR, namely those of the quadratic and quartic terms in the scalar potential. By integrating out the radial modes, one obtains a non-linear sigma model whose temperature, $T$, is the tuning parameter constructed from these couplings. The above steps produce a local fermionic theory with a non-zero bare mass and irrelevant interactions with a coupling constant $T$. We will show that for a range of bare fermion masses in the UV gauge theory there is a critical temperature $T_c$ where the interactions cancel the effects of the bare mass so that the dual fermion becomes massless, and this $T_c$ is within the regime of applicability of perturbation theory in $T$.\footnote{We emphasize that the appearance of the fixed point at a small value of $T$ is not a fortunate accident, but rather a consequence of the fact that we choose the bare fermion mass, whose magnitude is invisible in the IR, to be small compared to the lattice scale.} The UV cutoff provided by the lattice is quite useful in this respect, as it provides the scale that determines this regime. We can then study physics at an IR scale arbitrarily far below that of the UV, where the parameters of the lattice gauge theory's effective field theory will hardly appear perturbative and the bare fermion masses will hardly appear small. But, if we can identify the UV as describing a free massless fermion, then surely the same can be said for the IR. In short, performing a change of variables in the UV has a significant effect on the form of the renormalization flow, so that we can either obtain a strongly-coupled or free theory.

The outline of the rest of the paper is as follows. In section 2, we describe the lattice proof of the $U(N)$ dualities with $N_f=1$. In the following section, we repeat this analysis for the $SO(N)$ dualities. We then extend the construction to $N_f>1$.

As this work was nearing completion, we learned of the forthcoming work \cite{xu:so}, which has some overlap with section \ref{sec:so}.

\section{$N_f=1$ Free Dirac Fermion as Complex Boson Coupled to $U(N)_1$}

In this section we give an explicit lattice derivation of the $K=N_f=1$ case of \eqref{eq:Udual}, generalizing the $N=1$ construction in \cite{Chen:2017lkr}. The duality in Euclidean signature is explicitly \cite{Aharony:2015mjs, Hsin:2016blu}\footnote{The trace in the bosonic theory can be expanded as $\tra \parens{ bdb - \frac{2i}{3} b^3} + 2A \, d\tra b + N AdA$, using $b+A\equiv b+A\mathbbm{1}$.}
\begin{align}
-\mathcal{L}_{fermion} &= \bar\psi \gamma^\mu (\nabla_\mu - iA_\mu) \psi + m\bar\psi \psi + \frac{1}{2} \left( \frac{i}{4\pi} AdA + i \, 2\mathrm{CS_{grav}} \right) \nonumber \\[.1cm]
&\updownarrow \label{duality_UN} \\ 
-\mathcal{L}_{boson} &= -|(\nabla_\mu-ib_\mu) \phi |^2 - r |\phi|^2 - \frac{\lambda}{2}\left(|\phi|^2\right)^2 \nonumber \\[.1cm]
& \ \ \ \: + \frac{i}{4\pi} \tra\left((b+A)d(b+A)-\frac{2i}{3}(b+A)^3\right) + i\, 2N\mathrm{CS_{grav}} \ . \nonumber
\end{align}
Here $\psi$ is a Dirac fermion, $\phi$ a complex boson with $N$ colors, $b$ a $U(N)$ dynamical gauge field, and $A$ a background ``electromagnetic'' $\Spin_c$ connection. The level-$1/2$ CS term on the fermion side should be understood as coming from integrating out a heavy ``doubler'' Dirac fermion with $m\rightarrow -\infty$, or alternatively, as $+\pi\eta/2$ in terms of the eta-invariant \cite{Hsin:2016blu}. The duality is supposed to hold with $\sgn(r)=\sgn(m)$, and most interestingly at the critical point $r=m=0$.

In Euclidean signature we choose $\gamma^\mu$ to be the Pauli matrices $\sigma^\mu$ and treat $\psi$ and $\bar\psi$ as independent. This famously leads to a reflection positive, but not real, action.\footnote{In Euclidean signature, we can define a new notion of complex conjugation, $\psi^\dagger = i\bar\psi$, under which the massless Lagrangian $i\psi^\dagger\slashed\partial\psi$ is real. However, this should be regarded as a distraction, since the important condition for a Euclidean action is reflection positivity. Indeed, in the massive or Majorana cases the action cannot be made real.} Our conventions for Wick rotation to Minkowski signature are such that $\psi$ and $\bar\psi$ are invariant, while the coordinate $y$ becomes $it$, and correspondingly $\gamma^t = -i\sigma^y$. In Minkowski signature we also relate $\psi$ and $\bar\psi$ via $\bar\psi = -i\psi^\dagger \gamma^t = -\psi^\dagger\sigma^y$, so that the action is real.

\subsection{Lattice Constructions}

We will construct two lattice gauge theories representing the two sides of the duality and show that they are manifestly equivalent. We work on a cubic lattice representing the three-dimensional flat spacetime; we will discuss how to incorporate a gravitational background later. A lattice site is labeled by $n=(x, y, z)$, and the link between the sites $n$ and $n+\hat\mu \ (\hat\mu=\hat{x}, \hat{y}, \hat{z})$ is labeled by $n\mu$. The lattice unit length is set to $1$. On the lattice sites there live matter fields while on the links there live gauge fields. Specifically, the theories are as follows.

On the Dirac fermion side, at each site $n$ there is a pair of two-component Grassmann variables $(\psi_n)^\alpha$ and $(\bar\psi_n)_\alpha$, where $\alpha=\uparrow, \downarrow$ is the Dirac spinor index. On each link $n\mu$ there is the background electromagnetic gauge field $e^{iA_{n\mu}}$ and its conjugate $e^{-iA_{n\mu}}$. The partition function takes the form
\begin{align}
& Z^\psi[A] = \int \mathcal{D}\psi \mathcal{D}\bar\psi \ e^{-S_W^\psi [A]-S_{int}}, \ \ \ \ \mathcal{D}\psi \mathcal{D}\bar\psi \equiv \prod_n d^2\psi_n \: d^2\bar\psi_n, \nonumber \\
& -S_W^\psi [A] \equiv \sum_{n\mu} \left( \bar\psi_{n+\hat\mu} \frac{-\gamma^\mu-1}{2} e^{iA_{n\mu}} \psi_{n} + \bar\psi_{n} e^{-iA_{n\mu}} \frac{\gamma^\mu-1}{2} \psi_{n+\hat\mu} \right) + \sum_n M_\psi\bar\psi_n \psi_n.
\label{Z_psi}
\end{align}
The properties of Wilson's lattice fermion $S_W$ \cite{Wilson:1974sk, Wilson1977} are reviewed in Appendix \ref{sec:Wilson_fermion}; we are particularly interested in the vicinity $M_\psi\sim 3$ \cite{Chen:2017lkr}, where there is a continuum Dirac mode whose mass $m$ changes from negative to positive as $M_\psi$ increases across $3$, while the remaining ``doubler'' Dirac modes with masses at the lattice scale contribute a net level-1/2 CS term for the background $A$. We have also included some possible lattice scale interactions $S_{int}$, which are irrelevant in the continuum, up to some renormalization of the IR mass $m$ that we will take into account later.

On the boson side, we realize the $N$-color complex boson by a $U(N)$ non-linear sigma model in the fundamental representation. More precisely, at each site $n$ there is a $U(N)$ matrix $(V_n)^a_{\ b}$ where $a, b=1, \ldots, N$ is the color index. The non-linear sigma model boson variable is given by $\phi^a_n=(V_n)^a_{\ b} \, \xi^b$, where the ``reference'' column vector is
\begin{align}
\xi^b= \left[ \begin{array}{c} 1 \\ 0 \\ \vdots \\ 0\end{array} \right]. \label{eq:xi}
\end{align}
Besides the scalar, there is also a dynamical gauge field, which is realized by a $U(N)$ matrix $(U_{n\mu})^a_{\ b}=(e^{ib_{n\mu}})^a_{\ b}$ on each link $n\mu$. There is again the background electromagnetic gauge field $e^{\pm iA_{n\mu}}$. The gauge field $(b+A)$ has a CS term in the IR. While it is tricky to directly implement CS action at the lattice scale, to implement it in the IR, we can use a lattice fermion $\chi^a$ in the fundamental representation of $U(N)$, with $1<M_\chi<3$ \cite{Golterman:1992ub, Chen:2017lkr} (see Appendix \ref{sec:Wilson_fermion}). Piecing together these ingredients, the boson side of the duality's partition function is
\begin{align}
& Z[A] = \int \mathcal{D}U \ Z^\sigma[U] \ Z^\chi[U, A], \ \ \ \ \mathcal{D}U \equiv \prod_{n\mu} (dU_{n\mu})_{\mathrm{Haar}}, \nonumber \\
& \hspace{.0cm} Z^\sigma[U] = \int \mathcal{D}V \ e^{-S_\sigma[U]}, \ \ \ \ \mathcal{D}V \equiv \prod_{n} (dV_{n})_{\mathrm{Haar}}, \nonumber \\
& \hspace{1cm} -S_\sigma[U] \equiv \frac{1}{T} \sum_{n\mu} \left( \frac{\xi^\dagger V_{n+\hat\mu}^\dagger U_{n\mu} V_n \xi + \xi^\dagger V_n^\dagger U_{n\mu}^\dagger V_{n+\hat\mu} \xi}{2} - 1 \right), \nonumber \\
& \hspace{.0cm} Z^\chi[U, A] = \int \mathcal{D}\chi \mathcal{D}\bar\chi \ e^{-S_W^\chi [U, A]}, \ \ \ \ \mathcal{D}\chi \mathcal{D}\bar\chi \equiv \prod_n d^{2N}\chi_n \: d^{2N}\bar\chi_n, \nonumber \\
& \hspace{1cm}  -S_W^\chi [U, A] \equiv \sum_{n\mu} \left( \bar\chi_{n+\hat\mu} \frac{-\gamma^\mu-1}{2} e^{iA_{n\mu}} U_{n\mu} \chi_{n} + \bar\chi_{n} U_{n\mu}^\dagger e^{-iA_{n\mu}} \frac{\gamma^\mu-1}{2} \chi_{n+\hat\mu} \right) + \sum_n M_\chi\bar\chi_n \chi_n.
\label{Z_phi}
\end{align}
Note that the $U(N)$ variables are integrated with the Haar measure,\footnote{\label{ft:noMaxwell}One might worry that a different prescription is required, so that the gauge field for the central $U(1)\subset U(N)$ is `non-compact' \cite{polyakov:compactLattice} (in the sense that there is no potential for the dual photon \cite{polyakov:dualPhoton} -- i.e., the global $U(1)$ symmetry corresponding to $A$ under which monopole operators are charged is unbroken). However, because of the absence of the Maxwell term this distinction is immaterial. See also footnotes \ref{ft:gaugeField} and \ref{ft:compact_non-compact}.}
and the non-linear sigma model $S_\sigma$ is a direct generalization of the $U(1)$ XY model, with the ``temperature'' $T$ controlling the radius.

We note that in \eqref{Z_phi} one may include a Yang-Mills term for $U$. In Appendix \ref{sec:YM}, we discuss the consequences of doing so. In particular, we demonstrate that it changes neither our procedure nor our conclusions.

Our claim is that one can explicitly show
\begin{align}
Z[A] \ \propto \ Z^\psi[A]
\label{claim}
\end{align}
for any background $A$, with some overall proportionality constant independent of $A$. The two sides will involve some different $M_\chi$ and $M_\psi$, such that $M_\psi$ is a function of $M_\chi$ and $T$; the fermion side will also involve some irrelevant interactions $S_{int}$. Moreover, when $M_\chi$ implements level-1 CS, there is some critical value of $T$ such that $\psi$ has the desired massless Dirac mode in IR.

\subsection{Procedure}

Our plan is to integrate out the gauge field $U$ and discover that the boson $\phi$ binds with one color component of $\chi$ to make a new fermion $\psi$, while the remaining components of $\chi$ become invisible in the IR. As a first step, we single out one color by adopting the unitary gauge where $V_n$ is the identity matrix and $\phi_n = \xi$ (see \eqref{eq:xi}) for all $n$.\footnote{In \cite{Chen:2017lkr}, this gauge fixing step is avoided by a division by the volume of the gauge group in (3.2).} \footnote{This is an incomplete gauge choice, since any $U(N-1)$ gauge transformation that fixes $\xi$ preserves our gauge, but it will suffice for our purposes.} \footnote{All Faddeev-Popov determinants in this paper are trivial. This is clear from the fact that our gauge choice does not involve the gauge field or the fermion which remain in the path integral after our gauge fixing.} Thus, each link $n\mu$ ends up contributing
\begin{align}
\int dU_{n\mu} \ \exp\left( \frac{\xi^\dagger (U_{n\mu} +U^\dagger_{n\mu}) \xi - 2}{2T} + \bar\chi_{n+\hat\mu} \frac{-\gamma^\mu-1}{2} e^{iA_{n\mu}} U_{n\mu} \chi_{n} + \bar\chi_{n} U_{n\mu}^\dagger e^{-iA_{n\mu}} \frac{\gamma^\mu-1}{2} \chi_{n+\hat\mu} \right)
\end{align}
to $Z[A]$. As $U_{n\mu}$ does not appear elsewhere, the integral is done on each link separately \cite{Chen:2017lkr}. For definiteness, let's choose $\gamma^\mu=\sigma^\mu$ and look at a link $nz$ without loss of generality. The integral is
\begin{align}
\int dU_{nz} \ \exp\left( \frac{\xi^\dagger (U_{nz}+U^\dagger_{nz}) \xi - 2}{2T} - \bar\chi_{n+\hat{z}\, \uparrow} e^{iA_{nz}} U_{nz}  \chi_n^\uparrow - \bar\chi_{n\, \downarrow} U_{nz}^\dagger e^{-iA_{nz}} \chi_{n+\hat{z}}^\downarrow \right).
\label{Z_link}
\end{align}
Note that on each link, only one spinor component of each Grassmann variable appears.

To get an idea what will happen under the $U$ integral, let's consider the $T\rightarrow \infty$ and the $T\rightarrow 0$ limits. The $T\rightarrow \infty$ limit is equivalent to starting with $N_f=0$. One expects the strongly fluctuating $U$ to confine the $\chi$'s into massive color singlets that are invisible in the IR. In the integration \eqref{Z_link}, the exponent can be exactly expanded to finite order in the $4N$ Grassmann variables $\bar\chi_{n+\hat{z}\, \uparrow}^a$, $\bar\chi_{n\, \downarrow}^a$, $ \chi_n^{a,\uparrow}$, $\chi_{n+\hat{z}}^{a,\downarrow}$. These expanded terms form a polynomial in $U$ and $U^\dagger$. A term in this polynomial survives the $dU$ integral only if it has equal numbers of $U$ and $U^\dagger$ matrices. This in turn means the surviving terms must be independent of $A$, and must involve $4k \ (k=0, \ldots, N)$ Grassmann variables, forming color singlets on both sites $n$ and $n+\hat{z}$.\footnote{The result of the integration can be expressed in terms of \emph{Weingarten functions}, but we do not need the details here.} These terms involving $4k$ Grassmann variables can either be viewed as $2k$-body interactions across the link $nz$, or as the hopping of heavy color singlet bosonic objects, made out of $2k$ fermions, across the link $nz$.\footnote{There is no analytic proof that these order $1$ complicated terms will make the bosonic objects massive and invisible in the IR, but this is highly plausible on physical grounds, and is necessary for the duality to hold at $N_f=0$.} Thus, when $T\rightarrow \infty$ (or equivalently, at $N_f=0$) the theory is (almost) trivial\footnote{The $N_f=0$ theory is the $U(0)_1$ theory with a vanishing Lagrangian discussed in \cite{Hsin:2016blu}. Intuitively, the purpose of this theory is to preserve the memory that our theory once had fermions and required a spin structure until we coupled it to $A$. See also our discussion in section \ref{sec:so}.} in the IR. This agrees with the expectation from the IR theory \eqref{duality_UN} in the $r, m \rightarrow +\infty$ limit.

In the opposite $T\rightarrow 0$ limit, the integrand will be non-vanishing only if $U$ leaves $\xi$ invariant, i.e. the $U(N)$ gauge field $U^a_{\ b}$ is spontaneously broken to a $U(N-1)$ field ${U'}^{A}_{\ B}$ acting on the colors $B=2, \ldots, N$. Thus, \eqref{Z_link} becomes 
\begin{align}
& \exp\left( -\bar\psi_{n+\hat{z}\, \uparrow} e^{iA_{nz}} \psi_n^\uparrow - \bar\psi_{n\, \downarrow} e^{-iA_{nz}} \psi_{n+\hat{z}}^\downarrow \right) \nonumber \\[.2cm]
& \cdot \int dU'_{nz} \ \exp\left( -\bar\chi'_{n+\hat{z}\, \uparrow} e^{iA_{nz}} U'_{nz} {\chi'}_n^\uparrow - \bar\chi'_{n\, \downarrow} {U'}_{nz}^\dagger e^{-iA_{nz}} {\chi'}_{n+\hat{z}}^\downarrow \right)
\end{align}
where $\psi=\chi^{a=1} = \xi^\dagger \chi$ is the first color component of $\chi$, and $(\chi')^A = \chi^A$ are the remaining $N-1$ color components. Now $\psi$ is fully decoupled from $\chi'$ (the same is true in the mass term); in particular, $\psi$ is a free Wilson fermion with $M_\psi=M_\chi$. On the other hand, the $dU'$ integral involving the decoupled $\chi'$ degrees of freedom is the same as the above $dU$ integral in the $T\rightarrow \infty$ limit with $N$ replaced by $N-1$, and hence $\chi'$ is completely invisible in the IR. Thus, all we have is $Z^\psi$ with $M_\psi=M_\chi$ (and with $S_{int}$ fully decoupled from $\psi$). Since we have chosen $1<M_\chi<3$ to implement level-$1$ CS, $\psi$ will now implement a level-$1$ CS term for the background field $A$. This matches with the $r, m<0$ phase (since $m=M_\psi-3=M_\chi-3$, as explained in Appendix \ref{sec:Wilson_fermion}) from the IR theory \eqref{duality_UN}.

We are, in the end, interested in the finite $T$ case where an $m=0$ Dirac mode is developed in the IR. From the discussion above we expect $\chi^{a=1}=\xi^\dagger \chi$ on the boson side to become $\psi$ on the fermion side. Indeed, this has to happen because after the $dU$ integral, any term must be built out of color singlets on both sites $n$ and $n+\hat{z}$, and the only possible quadratic terms (in $\chi$) are $(\bar\chi_{n+\hat{z}\, \uparrow} \xi)(\xi^\dagger  \chi_n^\uparrow)$ and $(\bar\chi_{n\, \downarrow} \xi)(\xi^\dagger \chi_{n+\hat{z}\, \downarrow})$. In other words, from the UV perspective, $\chi^1$ is singled out by a Higgsed gauge field, while from the IR perspective, $\chi^1$ plays the role of the monopole operator binding with the boson $\phi$. What we still need to verify is that as $T$ increases from $0$, the IR mass of $\psi$ will increase from $m=M_\chi-3<0$ and hit $m=0$. Now there comes a nice aspect of the lattice gauge theory construction. We are free to set the IR energy scale arbitrarily low compared to the inverse lattice scale, so we can arrange the parameters such that
\begin{align}
\mbox{IR energy scale of interest} \ \ll \ |M_\chi-3| \ \ll \ 1 \ \equiv \ \mbox{Inverse lattice scale}.
\label{arrange_scales}
\end{align}
We have shown $m=M_\chi-3<0$ at $T=0$. Now that we have arranged $M_\chi$ very close to $3$, we expect a massless Dirac mode for $\psi$ will appear, if at all, at some finite but small $T_c \sim 3-M_\chi$. We can thus expand in $T$ and check that a small (compared to the inverse lattice scale) but non-zero $T$ indeed helps to increase $m$ so that it hits $0$ at some small $T=T_c$.\footnote{An alternative to our approach, where we make a small $T$ expansion before performing the integral  \eqref{Z_link}, might be available: one might be able to use the Itzytson-Zuber formalism \cite{Itzykson:1979fi, Balantekin:2000vn, ZinnJustin:2002pk}. However, the expansion is necessary for computing the IR mass $m$ anyways, as well as for making contact with the continuum \`a la footnote \ref{ft:gaugeField}, so we may as well employ it from the beginning; in fact, it helps to clarify the physics under consideration.} This low temperature expansion in the UV is fully under control, despite the strong coupling nature of the problem in the IR.

\subsection{Integrating out the Gauge Field}

To perform the low temperature expansion, it is natural to separate $U(N)$ into the $U(N-1)$ part that does not act on $\xi$ and the $U(N)/U(N-1)$ part that acts on $\xi$:
\begin{align}
U_{ab} \ = \ \exp \left(i\left[ \begin{array}{c|c} \theta & \ \ \ \ \eta_{C}^\ast \ \ \ \ \\[.11cm] \hline &\\ \eta_{A} & 0 \\ & \end{array} \right] \right) \ \cdot \ \left[ \begin{array}{c|c} 1 & \ \ \ \ 0 \ \ \ \ \\[.09cm] \hline &\\ 0 & U'_{C B} \\ & \end{array} \right]. \label{eq:thetaEta}
\end{align}
In this notation,
\begin{align}
\exp\left(\frac{\xi^\dagger (U+U^\dagger) \xi - 2}{2T}\right) = \exp\left(-\frac{\theta^2+|\eta|^2}{2T} + \frac{(\theta^2+|\eta|^2)^2+\theta^2 |\eta|^2}{24T} + \cdots \right).
\end{align}
Now we rescale $\theta$ and $\eta_{A}$ by $\sqrt{T}$, and due to the smallness of $T$, we can take the integration ranges of $\theta$ and $\eta_{A}$ to be $\mathbb{R}$ and $\mathbb{C}$ respectively; an overall constant from the Jacobian of this rescaling is omitted.\footnote{One might worry that we need a Jacobian in the change of variables from $U$ to $\{\theta,\eta,U'\}$. However, thanks to this rescaling, and the fact that the Lie algebra yields (via exponentiation) Riemann normal coordinates on the group manifold, the $(\theta,\eta)$-dependence in the Jacobian is $\Oo(T)$. As we will shortly explain in footnote \ref{ft:sneakiness}, this makes the Jacobian inconsequential. It is also important that the Jacobian does not yield terms odd in $\theta$ or $\eta$, since we drop terms that are odd in these variables.

(The statement about normal coordinates obtains after combining a few standard results (see, e.g., \S4 of \cite{taylor:lie}, chapter 18 of \cite{gallier:lie}, and \cite{milnor:lie}) about compact connected Lie groups. There is always a bi-invariant metric whose volume form is the Haar measure (which is also always bi-invariant). Indeed, when the Lie algebra is simple (e.g. $\mathfrak{su}(N)$ or $\mathfrak{so}(N)$), the Cartan-Killing form is the unique such metric, up to multiplication by a positive constant. For any bi-invariant metric, the geodesics starting at the identity are precisely the one-parameter groups, $e^{itX}$, where $t\in\mathbb{R}$ and $X$ is in the Lie algebra. Said another way, with this metric, the Lie group and Riemannian exponential maps coincide. Finally, since right multiplication is an isometry of this metric, the geodesics originating at a group element $U'$ are of the form $e^{itX}U'$.)} The integral \eqref{Z_link} on the link $nz$ can be expanded in powers of $T$ (we absorb $e^{iA_{nz}} U'_{nz} \rightarrow U'_{nz}$ and omit the $nz$ subscript common to all gauge fields):
\begin{align}
& \int dU' \int d\theta\, d^{2(N-1)} \eta \ \ \exp\left(-\frac{\theta^2+|\eta|^2}{2} \right) \ \left[ 1 + T\frac{(\theta^2+|\eta|^2)^2+\theta^2 |\eta|^2}{24} + \mathcal{O}(T^2) \right] \nonumber \\
& \hspace{.1cm} \exp\left( -\bar\psi_{n+\hat{z}\, \uparrow} e^{iA} \psi_n^\uparrow - \bar\psi_{n\, \downarrow} e^{-iA} \psi_{n+\hat{z}}^\downarrow -\bar\chi'_{n+\hat{z}\, \uparrow} U' {\chi'}_n^\uparrow - \bar\chi'_{n\, \downarrow} {U'}^\dagger {\chi'}_{n+\hat{z}}^\downarrow \right) \nonumber \\
& \hspace{.1cm} \left[ \ 1 \: + \frac{T}{2} \left(\bar\psi_{n+\hat{z}\, \uparrow} (\theta^2 + |\eta|^2) e^{iA} \psi_n^\uparrow + \bar\psi_{n\, \downarrow} e^{-iA} (\theta^2 + |\eta|^2) \psi_{n+\hat{z}}^\downarrow + \bar\chi'_{n+\hat{z}\, \uparrow} \eta \: \eta^\dagger U' {\chi'}_n^\uparrow + \bar\chi'_{n\, \downarrow} {U'}^\dagger \eta \: \eta^\dagger {\chi'}_{n+\hat{z}}^\downarrow \right) \right. \nonumber \\
& \hspace{.1cm}  \left. \ \ \ \ \ + \ T \left( \phantom{\frac{}{}\!\!} \left(\bar\psi_{n+\hat{z}\, \uparrow} \theta \psi_n^\uparrow \right) \left( \bar\psi_{n\, \downarrow} \theta \psi_{n+\hat{z}}^\downarrow\right) + \left(\bar\psi_{n+\hat{z}\, \uparrow} \: \eta^\dagger U' {\chi'}_n^\uparrow\right) \left( \bar\chi'_{n\, \downarrow} U'^\dagger \eta \: \psi_{n+\hat{z}}^\downarrow \right) \right. \right. \nonumber \\
& \hspace{.1cm} \left. \left. \hspace{6.3cm} + \left(\bar\chi'_{n+\hat{z}\, \uparrow} \eta \: \psi_n^\uparrow\right) \left( \bar\psi_{n\, \downarrow} \: \eta^\dagger {\chi'}_{n+\hat{z}}^\downarrow \right) \right) + \ \mathcal{O}(T^2) \phantom{\frac{1}{1}} \right] \ ,
\label{gauge_field_expansion}
\end{align}
where in the expansion we have omitted terms that are odd in $\theta$ or holomorphic / anti-holomorphic in $\eta_{A}$, as they vanish upon integration; terms with repeated Grassmann variables also vanish. Now we can perform the Gaussian integrals over $\theta$ and $\eta$; note that the $T/24$ term in the first line just produces an overall constant plus $\mathcal{O}(T^2)$ terms.\footnote{This statement relies on the following manipulation: $1+CT + DT+\Oo(T^2)=(1+CT)(1+DT+\Oo(T^2))$.\label{ft:sneakiness}} The result to order $T$ is
\begin{align}
& \int dU' \ \exp\left( -\bar\psi_{n+\hat{z}\, \uparrow} e^{iA} \psi_n^\uparrow - \bar\psi_{n\, \downarrow} e^{-iA} \psi_{n+\hat{z}}^\downarrow -\bar\chi'_{n+\hat{z}\, \uparrow} U' {\chi'}_n^\uparrow - \bar\chi'_{n\, \downarrow} {U'}^\dagger {\chi'}_{n+\hat{z}}^\downarrow \right) \nonumber \\
& \hspace{.1cm} \left[ \ 1 \: + T\left(N-\frac{1}{2}\right) \left(\bar\psi_{n+\hat{z}\, \uparrow} e^{iA} \psi_n^\uparrow + \bar\psi_{n\, \downarrow} e^{-iA} \psi_{n+\hat{z}}^\downarrow \right) + T\left(\bar\chi'_{n+\hat{z}\, \uparrow} U' {\chi'}_n^\uparrow + \bar\chi'_{n\, \downarrow} {U'}^\dagger {\chi'}_{n+\hat{z}}^\downarrow \right) \right. \nonumber \\
& \hspace{.1cm}  \left. \ \ \ \ \ + \ T \left(\bar\psi_{n+\hat{z}\, \uparrow} \psi_n^\uparrow \right) \left( \bar\psi_{n\, \downarrow} \psi_{n+\hat{z}}^\downarrow\right) + 2T\left(\left(\bar\psi_{n+\hat{z}\, \uparrow} \: {\chi'}_n^{\uparrow\, a'}\right) \left( \bar\chi'{}_{n\, \downarrow}^{a'} \psi_{n+\hat{z}}^\downarrow \right) + \left(\bar\chi'{}_{n+\hat{z}\, \uparrow}^{a'} \psi_n^\uparrow\right) \left( \bar\psi_{n\, \downarrow} \: {\chi'}_{n+\hat{z}}^{\downarrow\, a'} \right) \right) \phantom{\frac{1}{1}\!\!\!\!\!} \right]
\end{align}
up to overall constants. Now we can re-exponentiate these terms. The terms quadratic in $\psi$ receive a renormalization factor of $(1-T(N-1/2))$, while the terms quadratic in $\chi'$ receive a renormalization factor of $(1-T)$. We can remove these factors by a wavefunction renormalization:
\begin{align}
\sqrt{1-T(N-1/2)} \: \psi \ \rightarrow \ \psi, \ \ \ \ \ \ \sqrt{1-T} \: \chi' \ \rightarrow \ \chi'.
\label{wf_renormalization}
\end{align}
After this rescaling, we arrive at 
\begin{align}
& \int dU' \ \exp\left[ \: -\bar\psi_{n+\hat{z}\, \uparrow} e^{iA} \psi_n^\uparrow - \bar\psi_{n\, \downarrow} e^{-iA} \psi_{n+\hat{z}}^\downarrow -\bar\chi'_{n+\hat{z}\, \uparrow} U' {\chi'}_n^\uparrow - \bar\chi'_{n\, \downarrow} {U'}^\dagger {\chi'}_{n+\hat{z}}^\downarrow \right. \nonumber \\
& \hspace{.1cm} \left. + \ T \left(\bar\psi_{n+\hat{z}\, \uparrow} \psi_n^\uparrow \right) \left( \bar\psi_{n\, \downarrow} \psi_{n+\hat{z}}^\downarrow\right) + 2T\left(\left(\bar\psi_{n+\hat{z}\, \uparrow} \: {\chi'}_n^{\uparrow\, a'}\right) \left( \bar\chi'{}_{n\, \downarrow}^{a'} \psi_{n+\hat{z}}^\downarrow \right) + \left(\bar\chi'{}_{n+\hat{z}\, \uparrow}^{a'} \psi_n^\uparrow\right) \left( \bar\psi_{n\, \downarrow} \: {\chi'}_{n+\hat{z}}^{\downarrow\, a'} \right) \right) \right]
\end{align}
(plus $\mathcal{O}(T^2)$). The same idea clearly works for links in the $x$ and $y$ directions too, with $\uparrow, \downarrow$ replaced by the eigenvectors of $\sigma^x$ and $\sigma^y$ respectively.

The redefinition \eqref{wf_renormalization} changes the mass term in \eqref{Z_phi}:
\begin{align}
M_\chi \bar\chi_n \chi_n \ \rightarrow \ M_\psi \bar\psi_n \psi_n + M_{\chi'} \bar\chi'_n \chi'_n \ ,
\end{align}
where, to linear order in $T$,
\begin{align}
M_\psi = M_\chi \: (1+T(N-1/2)) \ ,
\label{mass_renormalization}
\end{align}
and $M_{\chi'} = M_\chi \: (1+T)$. Piecing together all the above, we arrive at the form of $Z^\psi[A]$ given in \eqref{Z_psi}, with $M_\psi$ given above and the interactions given by
\begin{align}
e^{-S_{int}} =\  \exp &\left[T \sum_{n\mu}\left(\bar\psi_{n+\hat{\mu}} \frac{-\gamma^\mu-1}{2} \psi_{n} \right) \left( \bar\psi_{n} \frac{\gamma^\mu-1}{2} \psi_{n+\hat{\mu}}\right) \right] \nonumber \\[.2cm]
& \times \int \mathcal{D}\chi \mathcal{D}\bar\chi \: \mathcal{D} U' \ e^{-S_W^{\chi'}[U']} \exp \left[ 2T \sum_{n\mu} \left( \left(\bar\psi_{n+\hat\mu} \frac{-\gamma^\mu-1}{2} \chi'{}_{n}^{a'}\right) \left( \bar\chi'{}_{n}^{a'} \frac{\gamma^\mu-1}{2} \psi_{n+\hat\mu} \right) \right. \right. \nonumber\\[.2cm]
&\hspace{5.8cm} \left. \left. + \: \left(\bar\chi'{}_{n+\hat\mu}^{a'} \frac{-\gamma^\mu-1}{2} \psi_{n}\right) \left( \bar\psi_{n} \frac{\gamma^\mu-1}{2} \chi'{}_{n+\hat\mu}^{a'} \right) \right) \right].
\end{align}
The first line is a self-interaction of $\psi$, while the remainder is an interaction of $\psi$ mediated by the $\chi'$ sector. It seems the latter is complicated. However, it only affects the $\psi$ sector at order $T^2$, and hence to order $T$ we can decouple the $\chi'$ sector and simply take
\begin{align}
-S_{int} = T \sum_{n\mu}\left(\bar\psi_{n+\hat{\mu}} \frac{-\gamma^\mu-1}{2} \psi_{n} \right) \left( \bar\psi_{n} \frac{\gamma^\mu-1}{2} \psi_{n+\hat{\mu}}\right).
\label{S_int}
\end{align}
The reason is the following. As we discussed in the $T\rightarrow 0$ case, thanks to confinement, $\int \mathcal{D}U' e^{-S_W^{\chi'}[U']}$ will yield terms with $4k$ ($k$ runs from $0$ to $N-1$) $\chi'$ fields across each link. On the other hand, each $T\bar\psi \psi \bar\chi' \chi'$ interaction involves only two $\chi'$ fields. Therefore, to connect the $\chi'$ sector to the $\psi$ sector, an even number of $T\bar\psi \psi \bar\chi' \chi'$ interactions must take place,\footnote{One caveat is that $\int \mathcal{D}U' e^{-S_W^{\chi'}[U']}$ inherits the quadratic mass term. But, this cannot couple the $\chi'$ sector to $T\bar\psi \psi \bar\chi' \chi'$, due to their spinor structures being orthogonal. The mass term is associated with a lattice site, and the spinor structure on a site is $\bar\chi'_{\uparrow}\chi'_{\uparrow} +\bar\chi'_{\downarrow}\chi'_{\downarrow}$. All other $\chi'$ terms are associated with a link, $n\mu$, such that on either site at the ends of that link, $\bar\chi'$ and $\chi'$ have opposite spins in the $\mu$ direction. So the spinor structure in the mass term is orthogonal to that in all other terms that are associated with links.} i.e. these contributions are $\mathcal{O}(T^2)$.

In summary, we have shown that $Z[A]$ given by \eqref{Z_phi} is, up to overall constants, equivalent to $Z^\psi[A]$ given by \eqref{Z_psi} after integrating out $U$ and $\chi'$. The lattice mass $M_\psi$ is given by \eqref{mass_renormalization} and the lattice scale interaction $S_{int}$ is given by \eqref{S_int}. This analysis is made to order $T$, which is controlled and sufficient, as we discussed below \eqref{arrange_scales}. Note that to this order, the only place $N$ appears is in \eqref{mass_renormalization}; for $N=1$, the above reduces to the $U(1)$ result \cite{Chen:2017lkr} as $I_0(1/T)/I_1(1/T) \rightarrow T/2$ at small $T$. At higher orders in $T$, the form of $Z^\psi[A]$ is unchanged, though $M_\psi$ and $S_{int}$ will receive higher order corrections.

Along the same lines of reasoning, one can also show the $2k$-point correlation functions satisfy
\begin{align}
&\left\langle \psi_{n_1} \cdots \psi_{n_k} \bar\psi_{\wt{n}_1} \cdots \bar\psi_{\wt{n}_k}  \right\rangle_{A} \nonumber \\[.2cm]
=& \ (1-T(N-1/2))^k \left\langle \left(\xi^\dagger V^\dagger_{n_1}\chi_{n_1}\right) \cdots \left(\xi^\dagger V^\dagger_{n_k}\chi_{n_k}\right) \left(\bar\chi_{\wt{n}_1} V_{\wt{n}_1}\phantom{^\dagger \!\!} \xi\right) \cdots \left(\bar\chi_{\wt{n}_k} V_{\wt{n}_k}\phantom{^\dagger \!\!} \xi\right)  \right\rangle_{A} \ ,
\end{align}
where the expectation values on the two sides are evaluated using theories \eqref{Z_psi} and \eqref{Z_phi}, respectively, with an arbitrary background $A$.

\subsection{Vanishing of the IR Dirac Mass at $T=T_c$}

Now we have a single fermion theory \eqref{Z_psi}, with lattice mass $M_\psi$ given by \eqref{mass_renormalization} and lattice scale self-interaction $S_{int}$ given by \eqref{S_int}. Were it not for the interaction $S_{int}$, this would be a free theory with a Dirac mode near $p_\mu=0$ with mass $m=M_\psi-3$ (in addition to Dirac modes at other points in the Brillouin zone with masses of order the lattice scale, as explained in Appendix \ref{sec:Wilson_fermion}); recall that we have chosen $0< 3-M_\chi \sim T \ll 1$ in \eqref{arrange_scales}, so to first order we have $m= (M_\chi-3) + 3T(N-1/2)$, and indeed there is a solution $0<T_c \ll 1$ to the equation $m=0$. However, it is not legitimate to ignore $S_{int}$ since it makes an order-$T$ contribution to the IR mass.

In fact, this is its only important effect, since it is irrelevant. (It is a UV realization of a current-current interaction. Note that our description of this interaction as irrelevant relies on our perturbative setup: $T\ll 1$.) Explicitly, the IR mass of the Dirac mode near $p_\mu=0$ is given by
\begin{align}
m = M_\psi - 3 + \Sigma(p=0) \ ,
\label{IRmass}
\end{align}
where $\Sigma$ is the self-energy of $\psi$ at $p_\mu=0$ to first order in $T$.\footnote{More precisely, for small $p$ one parametrizes $\Sigma(p) = \Sigma(0) + (1/Z-1) i\gamma^\mu p_\mu + \mathcal{O}(p^2)$, where $Z=1+\mathcal{O}(T)$ is the wavefunction renormalization. The IR mass should be $Z$ times the right-hand-side of \eqref{IRmass}. Fortunately, to compute $m$ to order $T$, it suffices to take $Z=1$. \label{wavefunction_renormalization_footnote}} It suffices to compute $\Sigma$ with only one $S_{int}$ insertion. The computation is the same as the $U(1)$ case \cite{Chen:2017lkr}; the details can be found in Appendix \ref{sec:diagram}. We find $\Sigma(p=0)\simeq 0.113\, T$, i.e.
\begin{align}
m = M_\chi - 3 + T(3N-3/2+0.113) \ ,
\label{IR_mass}
\end{align}
which, thanks to $0< 3-M_\chi \sim T \ll 1$, can hit $m=0$ for some $0<T_c \ll 1$ as desired. This completes the exact lattice derivation of the duality.

If one wants a theory of $\psi$ that is not only free in the IR, but also on the lattice, one can simply include a counter-term \eqref{S_int} for the $\chi$ theory \cite{Chen:2017lkr}. By similar reasoning as above, when the $\psi$ theory has a $m=0$ mode, the corresponding $\chi$ theory, with the $S_{int}$ self-energy, implements level-$1$ CS.

\subsection{Gravitational Background and Topology}

By now we have carried out the lattice construction of the duality \eqref{duality_UN} on an infinite cubic lattice, representing infinite flat spacetime. We now verify that this construction yields the correct behavior with a gravitational background, and even with a non-trivial topology \cite{Hsin:2016blu}. In fact, these properties are naturally integrated into our construction. Regarding gravity, one can readily see that the $\chi$ fermion we have in \eqref{Z_phi} indeed reproduces the right coefficient of $\mathrm{CS_{grav}}$ in \eqref{duality_UN}. As for topology, the CS (or BF) terms that can be consistently put on a $\Spin_c$ manifold \cite{Seiberg:2016rsg, Seiberg:2016gmd, Hsin:2016blu} can always be obtained from integrating out heavy fermions.

To incorporate curved spacetime and non-trivial topology, we introduce the metric and spin connection on the lattice using the method of \cite{Brower:2016vsl}; the lattice building blocks might no longer be cubes. This procedure does not interfere with our main step, integrating out $U$ on each individual link, in the establishment of the duality. Therefore, our UV analysis goes through without substantial change. In these more general spacetimes, it would be harder to extract the IR physics, compared to infinite flat spacetime. Nevertheless, since the field theory duality holds only in the infrared, we need only concern ourselves with curvature as small as the IR scale in \eqref{arrange_scales}, so that the only change in the IR interpretation is the change from flat to slightly curved spacetime.

A final issue is that in gauge theory, the overall normalization of the partition function might contain topological information about the spacetime \cite{Schwarz:1978cn, Witten:1988hf, w:sDualMod} if it cannot be presented as a product of local factors. In our derivation we dropped overall constants; now let's look closely at them. There are three sources of overall constants. The first is the gauge redundancy of $U$, much of which we have absorbed by setting $\phi=\xi$; the remaining redundancy and the Faddeev-Popov determinant yield a product of local factors.\footnote{The exception to this is that one should include a factor for each connected component of spacetime, since constant `gauge transformations' are actual symmetries.} The second is the overall constants we dropped in integrating out $U$ and rescaling $\psi, \chi'$; these constants are associated with the sites and links, i.e. they are already presented as products of local factors. The third is the decoupled $\chi'$ sector; since these fermions bind into heavy bosonic objects after the $U'$ integration, their contribution can also be viewed as a local term that contains no information about the topology.

\section{$N_f=1$ Free Majorana Fermion as Real Boson Coupled to $SO(N)_1$} \label{sec:so}

Now we turn to an explicit lattice construction of the ``$SO(N)_K$ + $N_f$ real bosons $\leftrightarrow$ $SO(K)_{-N+\frac{N_f}{2}}$ + $N_f$ Majorana fermions'' duality in the $N_f=K=1$ case (where, again, the fermion side is free). The procedure is very similar to the Dirac case, with some minor differences.

We briefly discuss a subtlety with Euclidean Majorana fermions (see, e.g., \S 2.2.1 of \cite{Witten:2015aba}). With a Lorentzian metric, Majorana fermions satisfy a reality condition, which in our conventions is $(\psi^\dagger)^T=\psi$. In Euclidean signature, such a condition may no longer be imposed, since $\psi$ is in the pseudoreal fundamental representation of $SU(2)\cong \Spin(3)$. That is, $\psi$ is a complex 2-component spinor (in the sense that it resides in a vector space with complex coefficients), as in the Dirac case. The difference from the Dirac case is that in the Lorentzian signature one may express the path integral (including the action) solely in terms of $\psi$, and this remains the case after Wick rotation. Indeed, the Euclidean action is that obtained from the Dirac case by replacing $\bar\psi \to -\psi^T \sigma^y$. We will therefore use the shorthand $\bar\psi$ for $-\psi^T\sigma^y$; however, it should be understood that we path integrate only over $\psi$, and not $\bar\psi$.\footnote{Readers may be familiar with a similar discussion involving Weyl fermions in four dimensions. However, there one treats $\psi$ and $\bar\psi$ as independent 2-component complex spinors, each of which is to be path-integrated over. $\psi$ transforms in the fundamental representation of the first $SU(2)$ factor in $\Spin(4)\cong SU(2)\times SU(2)$, while $\bar\psi$ is in the fundamental of the second factor.} Thus, just as in Lorentzian signature, the path integral for a free Euclidean Majorana fermion is the Pfaffian of the Dirac bilinear form (again, see \cite{Witten:2015aba}).

The IR Majorana duality in Euclidean spacetime can be presented as \cite{Metlitski:2016dht, Aharony:2016jvv}
\begin{align}
-\mathcal{L}_{fermion} &= \frac{1}{2}\: \bar\psi \gamma^\mu \nabla_\mu \psi + \frac{m}{2}\bar\psi \psi + \frac{i}{2} \, \mathrm{CS_{grav}} \nonumber \\[.1cm]
&\updownarrow \label{duality_SO} \\ 
-\mathcal{L}_{boson} &= -\frac{1}{2}\: \left((\nabla_\mu-ib_\mu) \phi \right)^2 - \frac{r}{2} \phi^2 - \frac{\lambda}{4}\left(\phi^2\right)^2 + \frac{i}{4\pi} \frac{1}{2} \tra \left(bdb -\frac{2i}{3}b^3\right) + i N\, \mathrm{CS_{grav}}. \nonumber
\end{align}
Here $\psi$ is a Majorana fermion, $\phi$ is a real boson with $N$ colors, and $b$ is an $SO(N)$ $(N\geq 3)$ dynamical gauge field. Again the duality is supposed to hold with $\sgn(r)=\sgn(m)$, and most interestingly at the critical point $r=m=0$.

We do not couple the theories to a background $\Spin_c$ connection, since doing so is impossible for a Majorana fermion. That is, Majorana fermions require a choice of spin structure. This manifests itself in the fact that our phases are governed by so-called `almost trivial' or `invertible' spin-TQFTs \cite{dijkgraaf:spinTQFT}, namely the $SO(n)_1$ theories discussed in \cite{Seiberg:2016rsg} which are dual to theories whose Lagrangians are given by $-\Li = - in \,\mathrm{CS_{grav}}$. The latter formulation allows us to define these theories for all $n\in\mathbb{Z}$, and we have $SO(-n)_1=SO(n)_{-1}$. In particular, the $m,r\to\infty$ phase is simply $SO(N)_1$ plus the $i N\, \mathrm{CS_{grav}}$ term, which yields $SO(0)_1$. (Despite appearances, even the $n=0$ theory is non-trivial and requires a choice of spin structure.) Similarly, when $m,r\to -\infty$, the gauge group is Higgsed to $SO(N-1)$, and the Chern-Simons terms together yield $SO(1)_{-1}$. The coefficients of the gravitational Chern-Simons terms in \eqref{duality_SO} have been chosen \cite{Aharony:2016jvv} so that the dual theories have the same framing anomaly \cite{Witten:1988hf}. As above, they arise naturally in our setup from integrating out massive fermions as we flow to the infrared.

The lattice construction is an obvious variant of \eqref{Z_psi} and \eqref{Z_phi}. (We will only do the construction on an infinite cubic lattice representing flat spacetime; the incorporation of a gravitational background is straightforward, as discussed in the Dirac case.) On the Majorana fermion side, at each site $n$ there is a two-component Grassmann variable $(\psi_n)^\alpha$, and we denote $(\bar\psi_n)_\alpha\equiv -\psi_n^\beta (\sigma^y)_{\beta\alpha}$. The partition function takes the form
\begin{align}
& Z^\psi = \int \mathcal{D}\psi \ e^{-S_W^\psi-S_{int}}, \ \ \ \ \mathcal{D}\psi \equiv \prod_n d^2\psi_n, \nonumber \\
& -S_W^\psi \equiv \sum_{n\mu} \frac{1}{2}\left( \bar\psi_{n+\hat\mu} \frac{-\gamma^\mu-1}{2} \psi_{n} + \bar\psi_{n} \frac{\gamma^\mu-1}{2} \psi_{n+\hat\mu} \right) + \sum_n \frac{M_\psi}{2}\bar\psi_n \psi_n \nonumber \\
& \phantom{-S_W^\psi } \ \, = \sum_{n\mu} \psi_{n+\hat\mu}^T \sigma^y \frac{\gamma^\mu+1}{2} \psi_{n} - \sum_n \frac{M_\psi}{2}\psi_n^T \sigma^y \psi_n \ ,
\label{Z_psiM}
\end{align}
and $S_{int}$ is again some irrelevant lattice scale interaction. The IR Majorana modes are straightforwardly deduced from the Dirac case.

On the boson side, we realize the $N$-color real boson by an $SO(N)$ non-linear sigma model in the vector representation. That is, at each site $n$ there is a $SO(N)$ matrix $(V_n)^a_{\ b}$ where $a, b=1, \ldots, N$ is the color index, and the scalar is given by $\phi^a = (V_n)^a_{\ b} \, \xi^b$, where the ``reference'' column vector $\xi^b$ is again the unit vector pointing in the $b=1$ direction. The dynamical gauge field is realized by an $SO(N)$ matrix $(O_{n\mu})^a_{\ b}=(e^{ib_{n\mu}})^a_{\ b}$ on each link $n\mu$. The partition function is
\begin{align}
& Z = \int \mathcal{D}O \ Z^\sigma[O] \ Z^\chi[O], \ \ \ \ \mathcal{D}O \equiv \prod_{n\mu} (dO_{n\mu})_{\mathrm{Haar}}, \nonumber \\
& \hspace{.0cm} Z^\sigma[O] = \int \mathcal{D}V \ e^{-S_\sigma[O]}, \ \ \ \ \mathcal{D}V \equiv \prod_{n} (dV_{n})_{\mathrm{Haar}}, \nonumber \\
& \hspace{1cm} -S_\sigma[O] \equiv \frac{1}{T} \sum_{n\mu} \left( \xi^T V_{n+\hat\mu}^T O_{n\mu} V_n \xi - 1 \right), \nonumber \\
& \hspace{.0cm} Z^\chi[O] = \int \mathcal{D}\chi \ e^{-S_W^\chi [O]}, \ \ \ \ \mathcal{D}\chi \equiv \prod_n d^{2N}\chi_n, \nonumber \\
& \hspace{1cm}  -S_W^\chi [O] \equiv \sum_{n\mu} \chi_{n+\hat\mu}^T \sigma^y \frac{\gamma^\mu+1}{2} O_{n\mu} \chi_{n} + \sum_n \frac{M_\chi}{2} \chi_n^T \sigma^y \chi_n \ .
\label{Z_phiO}
\end{align}
Again the CS term for $b$ is dynamically generated by a massive -- but now Majorana -- fermion $\chi^a$ with $1<M_\chi<3$.

Our goal is again to show
\begin{align}
Z\ \propto \ Z^\psi \ ,
\end{align}
and more generally
\begin{align}
\left\langle \psi_{n_1} \right. &\left. \cdots \psi_{n_k} \bar\psi_{\wt{n}_1} \cdots \bar\psi_{\wt{n}_k}  \right\rangle \nonumber \\[.2cm]
= & \ (const.)^k \left\langle \left(\xi^T V^T_{n_1}\chi_{n_1}\right) \cdots \left(\xi^T V^T_{n_k}\chi_{n_k}\right) \left(\bar\chi_{\wt{n}_1} V_{\wt{n}_1}\phantom{^\dagger \!\!} \xi\right) \cdots \left(\bar\chi_{\wt{n}_k} V_{\wt{n}_k}\phantom{^\dagger \!\!} \xi\right)  \right\rangle,
\label{Majorana_corr}
\end{align}
with the parameters arranged according to $0<3-M_\chi \sim T \ll 1$, and in particular at some critical value of $T$.

The derivation procedure is the same as in the Dirac case, but with a caveat to be explained soon. The first step is to exploit the $SO(N)$ gauge freedom to fix $V_n=1$ at all sites $n$. Then, in the theory $Z$, we look at each individual lattice link $n\mu$, which contributes the factor
\begin{align}
\int dO_{n\mu} \ \exp\left( \frac{\xi^T O_{n\mu} \xi - 1}{T} + \chi_{n+\hat\mu}^T \sigma^y \frac{\gamma^\mu-1}{2} O_{n\mu} \chi_{n} \right).
\label{link_integral}
\end{align}
Let's again discuss the $T\rightarrow \infty$ and $T=0$ limits, in which the mentioned caveat will appear. As $T\rightarrow\infty$, the first term above vanishes and the theory is essentially at $N_f=0$. We then exactly expand the exponent into a polynomial of Grassmann variables and perform the $dO_{n\mu}$ Haar integral. Previously, in the $U(N)$ Dirac case, only the terms with equal numbers of $U$ and $U^\dagger$ matrices survived the Haar integral. By contrast, thanks to the Majorana condition only $O$ appears now, and the only terms that survive the Haar integral do so because $O$ has determinant $1$:
\begin{align}
\int dO \: O^{a_1 b_1} \cdots O^{a_N b_N} \ \propto \ \epsilon_{a_1\cdots a_N} \epsilon_{b_1\cdots b_N} \ .
\end{align}
These surviving terms describe the hopping of a massive color singlet object $\epsilon_{a_1 \cdots a_N} \chi^{a_1} \cdots \chi^{a_N}$ which is invisible in the IR.\footnote{Again, there is no analytic proof that this object is massive and invisible in the IR, but this is highly plausible on physical grounds, and is necessary to make the duality hold at $N_f=0$. This is also related to the statement that the gap for $\mathbb{Z}_2\subset O(N)$ charged excitations does not close \cite{Metlitski:2016dht,Aharony:2016jvv}, as this object is $\mathbb{Z}_2$-odd.} Thus, the theory at $T\rightarrow \infty$ (or equivalently, at $N_f=0$) is (almost) trivial in the IR. In the opposite $T=0$ limit, the parts of $O$ that rotate $\xi$ are infinitely Higgsed, leaving the residual gauge field $SO(N-1)$. Then, as in the Dirac case, the $\chi^{a=1} = \xi^T\chi$ component is singled out as $\psi$ (with $M_\psi=M_\chi$), and the remaining components fully decouple from $\psi$ and bind into $SO(N-1)$ color singlets, which become invisible in the IR. This also explains why the Majorana duality holds for $N\geq 3$; for $N=2$, there is no residual $SO(N-1)$ gauge field, so this case must be treated separately. Fortunately, it is identical to the $U(1)$ case that we have studied. Thanks to our choice of $1<M_\chi<3$, this gapped phase has a level-1 $\mathrm{CS_{grav}}$ term.

As in the Dirac case, we shall arrange the scales according to \eqref{arrange_scales} and perform a small $T$ expansion to confirm the existence of a small, but finite, $T_c$. At small $T$, it is natural to separate $SO(N)$ into the $SO(N-1)$ part that does not rotate $\xi$ and the $SO(N)/SO(N-1)$ part that rotates $\xi$:
\begin{align}
O_{ab} \ = \ \exp \left(\left[ \begin{array}{c|c} 0 & \ \ \ \ -\eta_{C} \ \ \ \ \\[.11cm] \hline &\\ \eta_{A} & 0 \\ & \end{array} \right] \right) \ \cdot \ \left[ \begin{array}{c|c} 1 & \ \ \ \ 0 \ \ \ \ \\[.09cm] \hline &\\ 0 & (O')_{C B} \\ & \end{array} \right].
\end{align}
Since fluctuations of $\eta_{A}$ are suppressed by the smallness of $T$, we can rescale $\eta_{A}$ by $\sqrt{T}$ and extend each of its components' range of integration to $\mathbb{R}$. We perform the $\eta$ integral in \eqref{link_integral} and keep the result to linear order in $T$. Defining ${\chi'}^{A} \equiv \chi^{A}$, the result of integrating out $\eta_{n\mu}$ in \eqref{link_integral} is
\begin{align}
\exp\left(\left( 1 - T\frac{N-1}{2} \right) \psi_{n+\hat\mu}^T \sigma^y \frac{\gamma^\mu+1}{2} \psi_{n} \right) \ \int dO'_{n\mu} \ \exp\left( \left( 1 - \frac{T}{2} \right) \chi'{}^T_{n+\hat\mu} \sigma^y \frac{\gamma^\mu+1}{2} O'_{n\mu} \chi'_{n} \right)
\end{align}
(up to $\mathcal{O}(T^2)$ corrections). Note that to order $T$, the Majorana fermion $\psi$ is free (the current-current interaction of the Dirac case is disallowed by the Majorana condition) but has a wavefunction renormalization, while the $\chi'$ fermions are completely decoupled from $\psi$ and form massive $SO(N-1)$ singlets. One can rescale
\begin{align}
\sqrt{1-T(N-1)/2} \: \psi \ \rightarrow \ \psi \ ,
\end{align}
so that the hopping terms retain the usual normalization. This rescaling affects the mass term as
\begin{align}
M_\psi \equiv M_\chi \: (1+T(N-1)/2) \ . \label{eq:soMass}
\end{align}
Thus, we have shown that \eqref{Z_phiO}, after integrating out the gauge field, is equivalent to \eqref{Z_psiM}, with $M_\psi$ given above and $S_{int}$ negligible at order $T$. Since the $\psi$ theory is free at this order, we know there is an IR Majorana mode with mass $m=M_\psi-3$ -- this is simpler than the Dirac case \eqref{IRmass} where there is also a self-energy contribution to be considered. As we started with $M_\chi$ slightly below 3 (recall the arrangement of scales \eqref{arrange_scales}), there is some small, positive value of $T$ at which $M_\psi$ hits $3$. By a similar procedure, one can show \eqref{Majorana_corr}, where the constant is $1-T(N-1)/2$.

\section{$N_f>1$ -- Pushing the Flavor Bound}\label{sec:Nf}

\subsection{The generic case: $U(N)$ with $N_f\le N$ and $SO(N)$ with $N_f \le N-2$}

We now generalize our construction to larger values of $N_f$. For concreteness we restrict to a $U(N)$ gauge group, and comment on the small differences with the $SO(N)$ case at the end. A natural guess for the appropriate non-linear sigma model might be $N_f$ unit-length scalars. However, the condition $\phi_i^\dagger\phi_i=1$ (no sum over $i$) is not invariant under the desired $SU(N_f)$ global symmetry. Furthermore, after coupling to the $U(N)$ gauge field and fixing a unitary gauge, one is still left with continuous vacuum degeneracy. The $T=0$ phase therefore has massless scalars, and hardly resembles the gapped phase we expect from the duality. This degeneracy also suggests that there are additional $SU(N_f)$-invariant relevant deformations that we may add to the theory, on top of that parametrized by $T$. However, the desired IR fixed point has only one relevant $SU(N_f)$-invariant deformation \cite{Hsin:2016blu}, so it cannot be reached by slightly increasing $T$ from 0. These considerations all make it clear that imposing $\phi_i^\dagger\phi_i=1$ for each $i$ does not yield the desired non-linear sigma model.

To determine the correct condition to impose, recall that we motivated the non-linear sigma model in the introduction by integrating out the massive radial mode. Clearly, the `radial modes' in the current case depend on the potential. This is described by the following three relevant $SU(N_f)$-invariant terms that we may add to the free scalar Lagrangian \cite{Hsin:2016blu}:
\be r\phi_i^\dagger\phi_i + \frac{\lambda}{2}\parens{\phi_i^\dagger\phi_i}^2 + \rho \parens{\phi_i^\dagger\phi_j \phi_j^\dagger \phi_i } \label{eq:potential} \ .\ee
Focusing on the first two terms, we can, as in the introduction, eliminate them in favor of the condition $\phi_i^\dagger\phi_i = N_f$ (where now, of course, we are summing over $i$). The final term in the potential is
\be \rho \sum_i (\phi_i^\dagger\phi_i)^2 + \rho \sum_{i\not=j} |\phi_i^\dagger \phi_j|^2 \ .\ee 
Since $N_f\le N$, it is geometrically clear that this is minimized when $\phi_i^\dagger\phi_j=0$ for all $i\not=j$. A little more thought (or considering the saturation of the Cauchy-Schwarz inequality) then shows that the first term is minimized (subject to $\phi_i^\dagger\phi_i=N_f$) when $\phi_i^\dagger\phi_i = 1$ for all $i$. These conditions can be unified into the $SU(N_f)$-invariant constraint
\be \phi_i^\dagger\phi_j = \delta_{ij} \ .\ee
This yields the appropriate non-linear sigma model. Geometrically, the $N_f$ scalars form an orthonormal set of $N_f$ vectors in $\mathbb{C}^N$. The space parametrized by these scalars is known as a complex Stiefel manifold, $V_{N_f}(\mathbb{C}^N) \cong U(N) / U(N-N_f)$.

We may now trivially generalize our gauge choice from the earlier sections:
\be \phi_i^J = \delta^J_i \ ,\quad \phi_i^A = 0 \label{eq:Nfgauge} \ ,\ee
where $J=1,\ldots,N_f$ is a \emph{color} index, as is $A=N_f+1,\ldots,N$. Again thinking geometrically, we have chosen our $N_f$ orthonormal vectors to be the first $N_f$ vectors in the standard basis for $\mathbb{C}^N$. We emphasize the important point that gauge fixing has eliminated any vacuum degeneracy.

From here, our earlier steps generalize easily. Our gauge group is Higgsed to $U(N-N_f)$, and this confines $\chi'^A = \chi^A$. We are then left with $\psi^I$, for which integrating out the massive components of the gauge field yields order $T$ interactions. These may be seen, as above, to cancel away the mass of $N_f$ Dirac modes at some critical temperature, $T_c$. That is, \eqref{IR_mass} is generalized to\footnote{This follows from the generalization of \eqref{eq:thetaEta}: $\theta$ becomes a $N_f\times N_f$ Hermitian matrix (with $N_f^2$ real degrees of freedom) and $\eta$ becomes a $(N-N_f)\times N_f$ complex matrix (with $2N_f(N-N_f)$ real degrees of freedom). The coefficient $N-N_f/2$ in \eqref{eq:uMass} is the sum of these degrees of freedom, divided by $N_f$ because this is spread over the $N_f$ flavors. The self-energy is computed in Appendix \ref{sec:diagram}.}
\be m = M_\psi - 3 + \Sigma(p=0) = M_\chi - 3 + T \parens{ 3\, \parens{N-\frac{N_f}{2}} + 0.113 \, N_f } \ , \label{eq:uMass} \ee
which has a positive coefficient of $T$, so the massless fermions again obtain at some critical $T_c$.

We pause to note that our derivation is particularly trustworthy when $N_f=N$, as in this case there is no $\chi'$ that needs to confine. This is fortunate, since all of the dualities with $N_f<N$ may then be derived via mass deformations. A similar observation holds for the extreme $SO(N)$ cases discussed in sections \ref{sec:new1} and \ref{sec:new2}.

In the $SO(N)$ case, we simply replace the complex Stiefel manifold by a real one, $V_{N_f}(\mathbb{R}^N) \cong O(N)/O(N-N_f)$, which is the space of $N_f$ orthonormal vectors in $\mathbb{R}^N$. We again choose the gauge \eqref{eq:Nfgauge}, and our gauge group is Higgsed to $SO(N-N_f)$. Since $SO(1)$ is trivial, we find the requirement that $N - N_f \ge 2$. Otherwise, as we discuss below, the story changes. The IR mass is now
\be m = M_\psi - 3 + \Sigma(p=0) = M_\chi - 3 + \frac{T}{2} \parens{ 3\parens{ N - \frac{N_f + 1}{2} } + 0.113\, (N_f-1)} \ . \label{eq:soNfmass} \ee
Again, the coefficient of $T$ is positive.

\subsection{$SO(N)$ with $N_f = N-1$} \label{sec:new1}

\begin{figure}
\centering{
\includegraphics[width=0.67\textwidth]{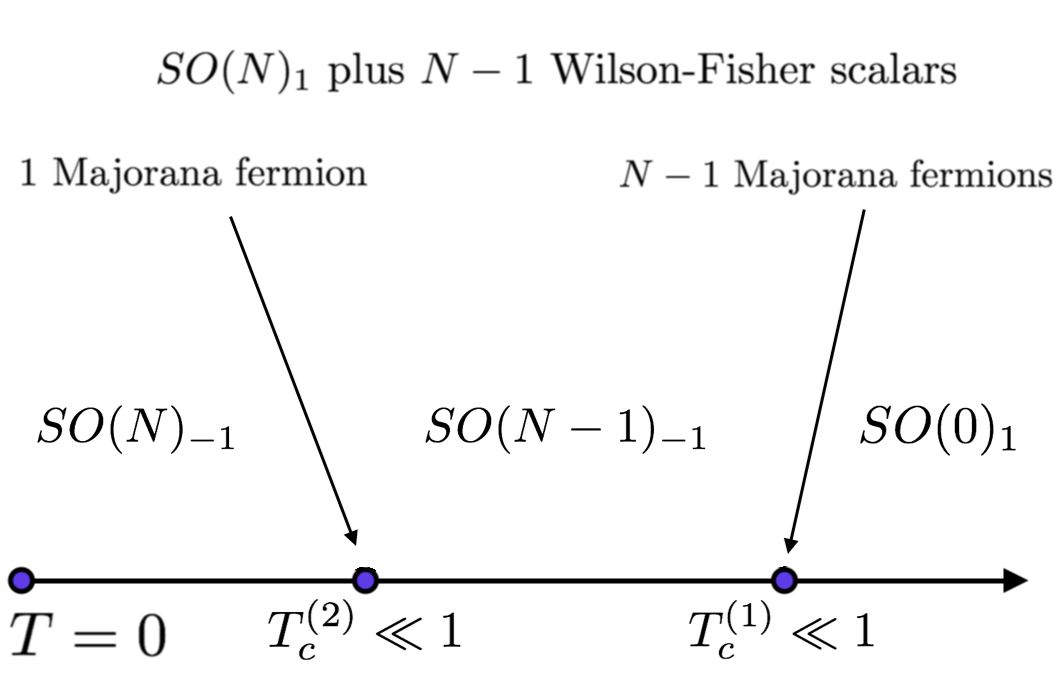}
\caption{Phase diagram of $SO(N)_1$ plus $N-1$ Wilson-Fisher scalars. Free fermionic descriptions suffice near the repulsive fixed points at $T_c^{(1)}$ and $T_c^{(2)}$, while the bosonic gauge theory applies at all $T$. Both fixed points are accessible in perturbation theory. The phases in this diagram are discussed below \eqref{duality_SO}. Away from the fixed points, a new relevant operator (associated to $m$ near $T_c^{(2)}$ or to $m'$ near $T_c^{(1)}$, or in the UV to $M_\chi$), which is invisible at the fixed points, drives the renormalization flow away from this line, so that there is no flow between the two fixed points.}\label{fig:nm1}}
\end{figure}

If $N-N_f=1$ (as in the $N=2,N_f=1$ case discussed above), then $\chi'$ is not confined, and we can find an extra light Majorana fermion in the dual theory. This was concretely observed when $N=2,N_f=1$, where we found a massless Dirac fermion instead of a Majorana one. However, that case turns out to be quite special, as only when $N=2$ are $\psi$ and $\chi'$ massless at the same value of $T$. This agrees with our CFT intuition, since in this case $\phi^\dagger\mathcal{M}$, where $\mathcal{M}$ is the monopole operator, being a Dirac fermion relies on the accident $SO(2)\cong U(1)$ that implies that the global symmetry acting on the monopole operator is $U(1)$, not $\mathbb{Z}_2$. More generally, the mass $m$ for $\psi^I$ and the mass $m'$ for $\chi'$ are respectively
\begin{align}
m &= M_\psi - 3 + \Sigma(p=0) = M_\chi - 3 + \frac{T}{2} \left( \frac{3}{2} + 0.113 \right) N \ , \label{eq:soNfmass1} \\[.2cm]
m' &= M_{\chi'} - 3 + \Sigma'(p=0) = M_\chi - 3 + T \left( \frac{3}{2} + 0.113 \right) (N-1) \ . \label{eq:chiPrimeMass}
\end{align}
(Note that $M_\psi$ in \eqref{eq:soNfmass1} is consistent with \eqref{eq:soNfmass}, but the self-energy is different; see Appendix \ref{sec:diagram}.) We find, for $N>2$, two different gauged Wilson-Fisher fixed points,\footnote{See Appendix \ref{sec:phaseTransitions} for a caveat regarding this terminology.} corresponding to two different critical temperatures. The situation is summarized in figure \ref{fig:nm1}. At
\be T_c^{(1)} = \frac{1}{3/2+0.113} \frac{3-M_\chi}{N/2} \ ,\ee
$\psi$ is massless, while at a lower temperature
\be T_c^{(2)} = \frac{1}{3/2+0.113}\frac{3-M_\chi}{N-1} \ ,\ee
$\chi'$ is massless. That is, we have the dualities
\be \mbox{$SO(N)_1$ plus $N-1$ Wilson-Fisher scalars at $T_c^{(1)}$ $\longleftrightarrow$ $N-1$ Majorana fermions} \ee
and
\be \mbox{$SO(N)_1$ plus $N-1$ Wilson-Fisher scalars at $T_c^{(2)}$ $\longleftrightarrow$ $1$ Majorana fermion} \ .\ee
Just as we can express the $N-1$ fermions in $\psi$ in a gauge-invariant manner as $\phi_i^T\chi$, we can also write $\chi'$ as
\be \chi' = \det([\phi\chi]) \equiv \epsilon^{a_1\cdots a_N}\phi_1^{a_1}\cdots \phi_{N-1}^{a_{N-1}}\chi^{a_N} \ .\ee
When $N=2$, which is equivalent to the $U(1)$ case, these fixed points coalesce and all $N=2$ Majorana fermions are massless at the same $T_c$.

The $N=2$ case is usually used as evidence that the usual dualities break down when $N-N_f = 1$, since starting from any such configuration one may flow (via Higgsing) to the $N=2,N_f=1$ case. We now see that while there are changes when $N-N_f = 1$, this is not the whole story, and indeed the fixed point at $T_c^{(1)}$ is quite similar to that of the generic duality. If we start with $N=3,N_f=2$ and Higgs away one color in order to study the $N=2,N_f=1$ theory, then we find the surprise that a new $U(1)$ symmetry emerges that guarantees that $\psi$ and $\chi'$ have the same mass. Acting with this symmetry on $\psi=\phi^T\chi$ yields an entire Dirac fermion, $\psi+i\chi'$. We are used to mass parameters mapping via the duality to mass parameters, but here that clearly cannot be the case, since there is only one mass parameter on the boson side while there are two Majorana masses available on the fermion side. The resolution of this is provided by noting that the latter mass terms are not invariant under the $U(1)$ global symmetry, and so they must map to monopole operators in the dual theory. Only the $U(1)$-invariant Dirac mass term maps to the scalar mass. Denoting the monopole operator by $\mathcal{M}$, we thus learn that
\be \mbox{$U(1)_1$ plus a Wilson-Fisher scalar and a $(\Real\phi^\dagger\M)^2$ potential $\longleftrightarrow$ 1 Majorana fermion} \ , \ee
as the potential on the left hand side is the Majorana mass $\bar\psi\psi$. 

One might now wonder if it is possible to have all $N$ Majorana fermions be simultaneously massless, when $N>2$, by allowing one of the scalars to have a temperature $t\not=T$. This would be extremely interesting, as it would mean that by breaking the $O(N-1)$ global symmetry in the bosonic gauge theory one could enhance the global symmetry in the dual theory to $O(N)$. However, it turns out that this is not possible. Instead, $\psi$ splits into $N-2$ fermions with a mass
\be M_\chi - 3 + \left( \frac{3}{2} + 0.113 \right) \left( (N-3) \frac{T}{2} + \frac{Tt}{T+t} + T \right) \label{eq:psi} \ee
and one fermion with a mass
\be M_\chi - 3 + \left( \frac{3}{2} + 0.113 \right) \left( (N-2) \frac{Tt}{T+t} + t \right) \ ,\ee
while $\chi'$ has a mass
\be M_\chi - 3 + \left( \frac{3}{2} + 0.113 \right) \left( (N-2) T + t \right) \label{eq:chiPrime} \ .\ee
For $N>2$, these masses can never be made equal for finite small $t\not=T$: the $\chi'$ mass is the greatest, while the first mass is greater / less than the second mass if $T$ is greater / less than $t$. Note that these formulae concretely demonstrate the symmetry enhancement described above as one Higgses from $N=3,N_f=2$ to $N=2,N_f=1$ by taking $t\to 0$: the masses \eqref{eq:psi} and \eqref{eq:chiPrime} adjust themselves in order to become equal.

\subsection{$SO(N)$ with $N_f = N$} \label{sec:new2}

When $N=N_f$, we are unable to choose the gauge \eqref{eq:Nfgauge}, since $SO(N)$ transformations cannot guarantee that our $N_f$ orthonormal vectors are oriented. So, there is a twofold vacuum degeneracy, labelled by the vevs \eqref{eq:Nfgauge} and the vev obtained by the replacement $\phi_1^1 = -1$. These vacua are related by the $\mathbb{Z}_2$ center of the $O(N_f)$ global symmetry group of the gauge theory. However, as we already remarked above, this $\mathbb{Z}_2$ should not be present at the IR fixed points we seek with free fermion duals.\footnote{This might seem strange, since the desired free fermionic dual will have $O(N_f)$ global symmetry. However, the $\mathbb{Z}_2$ present on the fermion side of the duality maps to a symmetry under which monopole operators of the gauge theory are charged \cite{Aharony:2016jvv}. More precisely, the operators that are odd under this $\mathbb{Z}_2$ are those monopole operators which are allowed in the $SO(N)$ gauge theory, but forbidden in the $\Spin(N)$ gauge theory.} So, we should have no qualms about employing spontaneous symmetry breaking in order to focus on the vacuum \eqref{eq:Nfgauge}. Indeed, we may as well break the $\mathbb{Z}_2$ symmetry explicitly: when $N=N_f$ we can include the potential
\be -\frac{1}{N!}\epsilon_{i_1\cdots i_N}\epsilon^{I_1\cdots I_N} \phi_{i_1}^{I_1}\cdots \phi_{i_N}^{I_N} = -\det \phi_i^I \label{eq:det} \ee
in our non-linear sigma model. For most values of $N$, this is dangerously irrelevant; i.e., it is irrelevant, but nevertheless important, as it dramatically affects the vacuum structure of the theory. In any case, since \eqref{eq:soNfmass} still holds (see Appendix \ref{sec:diagram} for the self-energy), we are lead, as above, to the following duality, again for $N\ge 2$:\footnote{Note that the $N=2$ case is different from $U(1)_1$ with $N_f=2$, since the quartic terms in the potential \eqref{eq:potential} are not independent, whereas the $SO(2)_1$ with $N_f=2$ theory has two independent quartic potential terms. The missing potential in the $U(1)$ case is of the form $-(\epsilon_{ij}\phi^\dagger_i\phi_j)^2$; it is forbidden by the $SU(2)$ flavor symmetry, but allowed by the $O(2)$ (or $SO(2)$, if we included \eqref{eq:det}) flavor symmetry of the $SO(2)$ theory, as is clear if we write this potential in terms of 2-component real scalars as $2(\epsilon_{ab}\phi_i^a\phi_j^b)^2 = 2((\phi_i^T\phi_i)^2 - (\phi_i^T\phi_j)^2)$. The last equality expresses this potential using the terms appearing in (the real scalar analogue of) \eqref{eq:potential}.}
\be \mbox{$SO(N)_1$ plus $N$ Wilson-Fisher scalars at $\tilde T_c^{(1)}$ $\longleftrightarrow$ $N$ Majorana fermions } \label{eq:NMajorana} \ .\ee
We emphasize that this last duality is qualitatively different from the rest discussed in this paper, due to the mechanism by which the renormalization group eliminates the $\mathbb{Z}_2$ symmetry from the infrared CFT. Foreseeing the existence of a second gauged Wilson-Fisher fixed point, as in the previous section, we have denoted the critical temperature of \eqref{eq:NMajorana} by $\tilde T_c^{(1)}$, which is again much smaller than the inverse lattice scale.

\begin{figure}
\centering{
\includegraphics[width=0.67\textwidth]{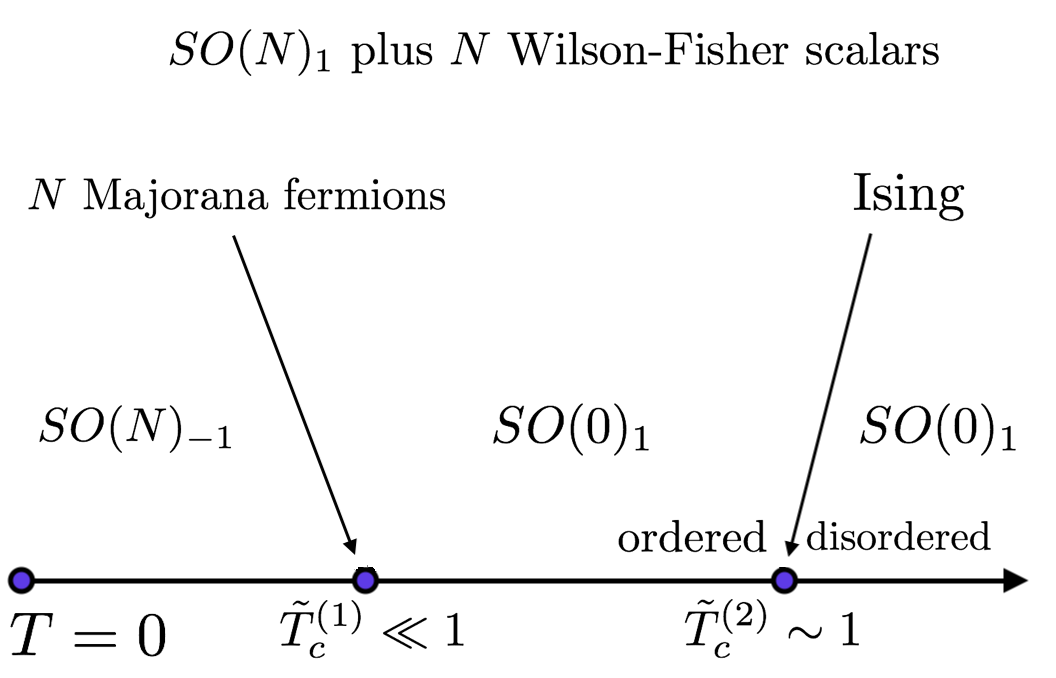}
\caption{Phase diagram of $SO(N)_1$ plus $N$ Wilson-Fisher scalars. Ungauged descriptions suffice near the repulsive fixed points at $\tilde T_c^{(1)}$ and $\tilde T_c^{(2)}$, while the bosonic gauge theory applies at all $T$. Only the free fermion fixed point is accessible in perturbation theory. The phases in this diagram are discussed below \eqref{duality_SO}. Away from the fixed points, a new relevant operator (associated to $M_\chi$), which is invisible at the fixed points, drives the renormalization flow away from this line, so that there is no flow between the two fixed points.}\label{fig:n}}
\end{figure}

The duality \eqref{eq:NMajorana}, of course, does not exist when $N=1$. However, we can now increase the temperature and search for a $\mathbb{Z}_2$ symmetry-restoring phase transition at some $\tilde T_c^{(2)} \sim \mathcal{O}(1) \gg \tilde T_c^{(1)}$, analogous to that described by the $N=1$ Ising fixed point. In fact, we will now argue that the Ising fixed point obtains for all $N$:
\be \mbox{$SO(N)_1$ plus $N$ Wilson-Fisher scalars at $\tilde T_c^{(2)}$ $\longleftrightarrow$ Ising } \label{eq:Ising} \ .\ee
This follows from the observation that the fermions' masses increase as we increase the temperature from the fixed point in \eqref{eq:NMajorana}. The minimal assumption is then that strong interactions do not yield new light degrees of freedom. We are thus led to the proposal of figure \ref{fig:n}.

Alternatively, if one is willing to believe that the gauge theory with an infinite Maxwell coupling is in the same universality class as the associated continuum theory even at finite $T$ (c.f. footnote \ref{ft:gaugeField}), then our usual arguments can provide additional evidence for the duality, as we now demonstrate for $N=2$. We fix the gauge $\phi_1^a=(s,0)^T$, $\phi_2^a=(0,1)^T$, where $s=\pm 1$ is an Ising variable. By regarding $SO(2)$ as $U(1)$, we can recast these as 1-component complex scalars -- $\phi_1=is$, $\phi_2=1$ -- which we succinctly write in the 2-component form $\phi=(is,1)^T$. We similarly regard $\chi$ as a Dirac fermion. Next, we exactly integrate out the $U(1)$ gauge field, as in \cite{Chen:2017lkr}; we will not use a small $T$ expansion since we expect $\tilde T_c^{(2)}$ to occur at order $1$. The contribution to the partition function from a link $n\mu$ is
\begin{align}
& \int_{-\pi}^\pi \frac{db_{n\mu}}{2\pi} \ \exp\left( \frac{\phi_{n+\hat\mu}^\dagger e^{ib_{n_\mu}} \phi_n + \phi_n^\dagger e^{-ib_{n\mu}} \phi_{n+\hat\mu} - 4}{2T}\right) \nonumber \\[.2cm]
& \ \ \ \ \ \exp\left( \bar\chi_n \frac{\gamma^\mu-1}{2} e^{ib_{n\mu}} \chi_{n+\hat\mu} + \bar\chi_{n+\hat\mu} e^{-ib_{n\mu}} \frac{-\gamma^\mu-1}{2} \chi_n \right) \ .
\end{align}
Note that the first factor is the exponential of $((1+s_n s_{n+\hat\mu}) \cos b_{n\mu} - 2 )/T$. We then Fourier expand the first exponential and Taylor expand the second. When $s_n s_{n+\hat\mu}=1$, we have
\begin{align}
& \int_{-\pi}^\pi \frac{db_{n\mu}}{2\pi} \ \sum_{j_{n\mu} \in \mathbb{Z}} e^{-2/T} I_{j_{n\mu}}(2/T) \ e^{ib_{n\mu} j_{n\mu}} \nonumber \\[.2cm]
& \ \ \left[ \: 1 + \bar\chi_n \frac{\gamma^\mu-1}{2} e^{ib_{n\mu}} \chi_{n+\hat\mu} + \bar\chi_{n+\hat\mu} e^{-ib_{n\mu}} \frac{-\gamma^\mu-1}{2} \chi_n + \left( \bar\chi_n \frac{\gamma^\mu-1}{2} \chi_{n+\hat\mu} \right) \left( \bar\chi_{n+\hat\mu} \frac{-\gamma^\mu-1}{2} \chi_n \right) \right] \nonumber \\[.3cm]
=& \  e^{-2/T} I_{0}(2/T) \ \exp\left[ \: \frac{I_1(2/T)}{I_0(2/T)} \left(\bar\chi_n \frac{\gamma^\mu-1}{2} \chi_{n+\hat\mu} + \bar\chi_{n+\hat\mu} \frac{-\gamma^\mu-1}{2} \chi_n  \right) \right. \nonumber\\[.2cm]
& \hspace{3.6cm} \left. + \left( 1-  \frac{I_1(2/T)^2}{I_0(2/T)^2}\right) \left( \bar\chi_n \frac{\gamma^\mu-1}{2} \chi_{n+\hat\mu} \right) \left( \bar\chi_{n+\hat\mu} \frac{-\gamma^\mu-1}{2} \chi_n \right) \: \right] \ ,
\end{align}
which is identical to the $N_f=1$ result of \cite{Chen:2017lkr} with the replacement $1/T\to 2/T$; recall $I_j=I_{-j}$ is the $j$th modified Bessel function. On the other hand, when $s_n s_{n+\hat\mu}=-1$, the contribution is simply
\begin{align}
& e^{-2/T} \left[ 1 + \left( \bar\chi_n \frac{\gamma^\mu-1}{2} \chi_{n+\hat\mu} \right) \left( \bar\chi_{n+\hat\mu} \frac{-\gamma^\mu-1}{2} \chi_n \right) \right] \nonumber \\[.2cm]
=& \ e^{-2/T} \ \exp\left[ \left( \bar\chi_n \frac{\gamma^\mu-1}{2} \chi_{n+\hat\mu} \right) \left( \bar\chi_{n+\hat\mu} \frac{-\gamma^\mu-1}{2} \chi_n \right) \right].
\end{align}
There are two main differences between $s_n s_{n+\hat\mu}=\pm 1$. The first difference is the overall factor of $I_0(2/T)$ that favors $\mathbb{Z}_2$ symmetry breaking at small temperatures. The second is that the fermion cannot hop through a link with $s_n s_{n+\hat\mu}=- 1$.

Now the phases can be easily understood. $I_0(2/T)$ behaves as $e^{2/T} \sqrt{T/4\pi}$ at small $T$ and $1+1/T^2$ at large $T$. Thus, at $T\ll 1$ only the $s_n s_{n+\hat\mu}=1$ configurations will be realized, and the theory becomes the same as $N_f=1$ except $T\rightarrow T/2$. In particular, at $\tilde T_c^{(1)}$ (where \eqref{eq:soNfmass} with $N=N_f=2$ vanishes, or equivalently where \eqref{eq:uMass} with $N=N_f=1$ and $T\rightarrow T/2$ vanishes), one finds a free Dirac fermion. As $T$ increases, the mass of this fermion increases. If we always had $s_ns_{n+\hat\mu}=1$, then we would suspect that this fermion remains massive at order-1 values of $T$, so if we want to search for dramatic effects that it causes we should study Ising domain boundaries on which $s_n s_{n+\hat\mu}=-1$. However, the fermion's correlation length is of order the lattice scale within each Ising domain, and it cannot hop through domain boundaries, so it seems unlikely to affect the Ising spin in any interesting way. Thus, we are left only with the Ising model at temperatures of order-1 and above. It is not hard to see the same physical picture carries over to higher values of $N$.

\section*{Acknowledgments}
M.Z. thanks L. Delacretaz, I. Esterlis, S. Kachru, S. Nezami, and S. Raghu for enjoyable conversations, and S. Kachru for his enduring support and positivity. We thank O. Aharony for comments on an early version of the manuscript, which motivated us to include the discussion in Appendix \ref{sec:YM}. J.-Y. C. is supported by the Gordon and Betty Moore Foundation's EPiQS Initiative through Grant GBMF4302.

\appendix

\section{Wilson's Lattice Fermion} \label{sec:Wilson_fermion}

We remind the reader of Wilson's lattice fermion action $S_W$ \cite{Wilson:1974sk, Wilson1977}. In the absence of a gauge field, the action in momentum space is
\begin{align}
-S_W = \int_{-\pi}^\pi \frac{d^3 p}{(2\pi)^3} \ \bar\psi_{-p} \left(\sum_\mu \gamma^\mu \, i\sin p_\mu + \left( M - R \sum_\mu \cos p_\mu \right)\right) \psi_p \ ,
\end{align}
where in the main text we have set $R=+1$. Consistent with the fermion doubling theorem \cite{Nielsen:1980rz}, there are $2^D=8$ IR Dirac modes, corresponding to the vicinities $p=\bar{p}+\delta p$ of the eight saddle points where each $\bar{p}_\mu$ component is either $0$ or $\pi$. Taking $\sin p_\mu \simeq \delta p_\mu \, \cos\bar{p}_\mu$ and $\cos p_\mu \simeq \cos\bar{p}_\mu$ around each of these saddle points, we see that each IR Dirac mode has mass $\pm(M-R\sum_\mu \cos \bar{p}_\mu)$ , where the sign in this expression is $\prod_\mu \cos\bar{p}_\mu$. The role of $R$ is therefore to make the modes have different masses \cite{Wilson:1974sk, Wilson1977}. For example, near $M\sim 3R$, the Dirac mode at $\bar{p}=0$ has a small IR mass $m=M-3R$, while the other seven Dirac modes have lattice scale masses.

When coupled to a slowly varying $U(1)$ gauge field $A_{n\mu}$, integrating out the fermion will produce a level-$C$ CS term, where each IR Dirac mode contributes $(-1/2)\, \sgn({\rm mass})$ to $C$. Therefore \cite{Golterman:1992ub}, 
\begin{align}
C= \left\{ \begin{array}{ll} 0, & \ \ \ \ 3|R|<|M| \\ \sgn(R), & \ \ \ \ |R|< |M|<3|R| \\ -2\, \sgn(R), & \ \ \ \ |M|<|R| \end{array} \right. \ .
\end{align}
If $M=3R$ exactly, then there will be one massless Dirac mode, and integrating out the remaining Dirac modes contributes $C=\sgn(R)/2$ (the IR meaning of which is supplied by our UV lattice regularization). Note that the magnitude of $R$ does not affect any IR physics (as long as we scale $M$ correspondingly), so we set $|R|=1$ in the main text, which has the UV convenience that $\pm \gamma^\mu-R$ projects out one spinor component.

If we replace $\psi$ by the $N$-color fermion $\chi^a$, the analysis is very much the same; in particular the non-Abelian CS level is still as above. Therefore, to implement level-$1$ CS, we can set $R=+1$ and $1<M_\chi<3$. A similar discussion holds for Majorana fermions and $SO(N)$ gauge fields.

\section{Mass Renormalization from Lattice Scale Interaction} \label{sec:diagram}

We compute the self-energy in \eqref{IRmass} and \eqref{eq:uMass} for $1\leq N_f \leq N$ Dirac fermions. We set the external momentum $p_\mu=0$, and take $M_\chi \simeq 3$ in the internal lines. The self-energy at $p=0$ is proportional to the identity matrix in $2\times 2$ spinor space, due to the charge conjugation symmetry $\psi \rightarrow \sigma^2 \bar\psi^T, \ \bar\psi \rightarrow -\psi^T \sigma^2$. To first order,
\begin{eqnarray*}
-\Sigma \ \mathbf{1}_{2\times 2} \ \delta_{IJ} \ \ = \ \ 
\parbox{33mm}{
\begin{fmffile}{zzz-Sigma0}
\begin{fmfgraph*}(32, 20)
\fmfleftn{l}{4}\fmfrightn{r}{4}
\fmf{fermion,label.side=left,label=$I$}{m,l2}
\fmf{fermion,label.side=left,label=$J$}{r2,m}
\fmffreeze
\fmf{phantom,tension=5}{m1,l4}
\fmf{phantom,tension=5}{m1,r4}
\fmf{phantom}{m1,m}
\fmf{photon,left=.9}{m,m1}
\fmf{photon,left=.9}{m1,m}
\fmfdot{m}
\end{fmfgraph*}
\end{fmffile}
}
\ + \ \
\parbox{33mm}{
\begin{fmffile}{zzz-Sigma00}
\begin{fmfgraph*}(32, 20)
\fmfleftn{l}{4}\fmfrightn{r}{4}
\fmf{fermion,label.side=left,label=$I$}{m,l2}
\fmf{fermion,label.side=left,label=$J$}{r2,m}
\fmffreeze
\fmf{phantom,tension=5}{m1,l4}
\fmf{phantom,tension=5}{m1,r4}
\fmf{phantom}{m1,m}
\fmf{dbl_wiggly,left=.9}{m,m1}
\fmf{dbl_wiggly,left=.9}{m1,m}
\fmfdot{m}
\end{fmfgraph*}
\end{fmffile}
}
\ + \ \ \ \
\parbox{33mm}{
\begin{fmffile}{zzz-Sigma2}
\begin{fmfgraph*}(32, 20)
\fmfleftn{l}{4}\fmfrightn{r}{4}
\fmf{fermion,label.side=left,label=$I$}{m1,l2}
\fmf{fermion,tension=0.5}{m2,m1}
\fmf{fermion,label.side=left,label=$J$}{r2,m2}
\fmffreeze
\fmf{photon,left=1}{m1,m2}
\fmfdot{m1,m2}
\end{fmfgraph*}
\end{fmffile}
}
\\[.0cm]
\,
\end{eqnarray*}
where the arrowed lines are the $\psi^I \: (I=1, \ldots, N_f)$ Dirac fermions, the wavy lines are the Hermitian $\theta_{IJ}$ (Higgsed) gauge fields, and the double wavy lines are the complex $\eta_{AJ}$ (Higgsed gauge fields). The $\bar\psi\psi \theta\theta$ and $\bar\psi\psi \eta^\dagger \eta$ vertices appeared in \eqref{gauge_field_expansion} (and its $N_f>1$ generalizations), and are responsible for renormalizing $M_\chi$ to $M_\psi$, which we have already taken into account. So we only need to compute the third diagram, whose  $\bar\psi \psi \theta$ vertices appeared in the quartic terms in \eqref{gauge_field_expansion}. (One can also draw an additional tadpole diagram, which vanishes by charge conjugation symmetry.) The Feynman rules are given by
\begin{eqnarray*}
\parbox{28mm}{
\begin{fmffile}{zzz-G}
\begin{fmfgraph*}(25, 10)
\fmfleftn{l}{3}\fmfrightn{r}{3}
\fmf{fermion,label.side=right,label=$k$}{r2,l2}
\fmf{phantom,label.side=right,label=$I$}{l2,r2}
\fmfdot{l2,r2}
\end{fmfgraph*}
\end{fmffile}
}
= \ \left[-\sum_\mu\left(\gamma^\mu i\sin k_\mu - \cos k_\mu \right) - M_\chi\right]^{-1} = \ \frac{-M_\chi+\sum_\mu \left(\cos k_\mu + \gamma^\mu \: i\sin k_\mu \right) }{(M_\chi-\sum_\nu \cos k_\nu)^2 + \sum_\nu (\sin k_\nu)^2}
\end{eqnarray*}
\vspace{.3cm}
\begin{eqnarray*}
\hspace{1.8cm}
\parbox{39mm}{
\begin{fmffile}{zzz-V}
\begin{fmfgraph*}(25, 25)
\fmfleftn{l}{5}\fmfrightn{r}{3}
\fmf{fermion,label.side=right,label=$J$}{l1,m}
\fmf{fermion,label.side=right,label=$I$}{m,l5}
\fmf{photon,label.side=right,label=$q$}{r2,m}
\fmflabel{$k-q/2$}{l2}
\fmflabel{$k+q/2$}{l4}
\fmflabel{$\mu, IJ$}{r2}
\fmfdot{m}
\momentumarrow{a}{up}{4}{r2,m}
\end{fmfgraph*}
\end{fmffile}
}
= \ -e^{-iq_\mu/2} \left(\gamma^\mu i\cos k_\mu + \sin k_\mu \right)
\end{eqnarray*}
\vspace{.3cm}
\begin{eqnarray*}
\hspace{1cm}
\parbox{36mm}{
\begin{fmffile}{zzz-U}
\begin{fmfgraph*}(20, 10)
\fmfleftn{l}{3}\fmfrightn{r}{3}
\fmf{photon,label.side=left,label=$q \ \mathrm{mod} \: 2\pi$}{l2,r2}
\fmfdot{l2,r2}
\fmflabel{$\mu, IJ \ $}{l2}
\fmflabel{$\ \nu, J'I'$}{r2}
\momentumarrow{a}{up}{4}{r2,l2}
\end{fmfgraph*}
\end{fmffile}
}
= \ T \: \delta_{\mu\nu} \: \delta_{II'} \delta_{JJ'}
\end{eqnarray*}
in Euclidean signature. (The vertex is given by $\partial_{k_\mu}$ of the inverse propagator due to gauge invariance, and there is an additional factor accounting for an Umklapp process.\footnote{At the vertex, we take $-\pi<k_\mu\pm q_\mu/2 \leq \pi$. But this implies $-2\pi<q_\mu< 2\pi$, i.e. $q_\mu$ would have a range of $4\pi$, so we allow it to change by $2\pi$ across the interaction line, which corresponds to an Umklapp process. If this happens, it gives rise to an extra factor $e^{-iq_\mu/2} e^{-i(\pm 2\pi - q_\mu)/2}=-1$. This issue does not come up in our self-energy computation.})
The third diagram is given by
\begin{align}
& \delta_{IJ} \, \delta_{KK} \, T \sum_\mu \int_{-\pi}^\pi \frac{d^3 k}{(2\pi)^3} \nonumber\\[-.1cm]
& \hspace{2.0cm} \left(\gamma^\mu i\cos \frac{k_\mu}{2} + \sin \frac{k_\mu}{2} \right) \ \frac{-M_\chi+\sum_\lambda \left(\cos k_\lambda + \gamma^\lambda \: i\sin k_\lambda \right) }{(M_\chi-\sum_\kappa \cos k_\kappa)^2 + \sum_\kappa (\sin k_\kappa)^2} \ \left(\gamma^\mu i\cos \frac{k_\mu}{2} + \sin \frac{k_\mu}{2} \right) \nonumber \\[.3cm]
= \ & \delta_{IJ} \, N_f \, T \int_{-\pi}^\pi \frac{d^3 k}{(2\pi)^3} \ \sum_\mu \frac{\cos k_\mu \: (M_\chi-\sum_\lambda \cos k_\lambda) - (\sin k_\mu)^2}{(M_\chi-\sum_\kappa \cos k_\kappa)^2 + \sum_\kappa (\sin k_\kappa)^2} \nonumber \\
= \ & \delta_{IJ} \, N_f \, T \int_{-\pi}^\pi \frac{d^3 k}{(2\pi)^3} \ \brackets{ -1 + \frac{M_\chi ( M_\chi - \sum_\lambda \cos k_\lambda)}{(M_\chi - \sum_\kappa \cos k_\kappa)^2 + \sum_\kappa (\sin k_\kappa)^2 } } \ .
\label{self-energy_calculation}
\end{align}
To first order we can take $M_\chi \simeq 3$. Performing the integration numerically, we find $-\Sigma(p=0) \approx -0.113 \, N_f \, T$. (One can also explicitly check that the first two diagrams renormalize $M_\chi$ to $M_\psi$ while the tadpole diagram vanishes.)

Next we consider the Majorana cases with $1\leq N_f \leq N-2$ and $N_f=N$; for the $N_f=N-1$ case one has to also take the unconfined $\chi'$ into account, as we will discuss later. The Feynman rules are
\begin{eqnarray*}
\parbox{30mm}{
\begin{fmffile}{zzz-G_M}
\begin{fmfgraph*}(25, 10)
\fmfleftn{l}{3}\fmfrightn{r}{3}
\fmf{plain,label.side=right,label=$k$}{r2,l2}
\momentumarrow{a}{up}{4}{r2,l2}
\fmf{phantom,label.side=right,label=$I$}{l2,r2}
\fmfdot{l2,r2}
\end{fmfgraph*}
\end{fmffile}
}
= \ \frac{-M_\chi+\sum_\mu \left(\cos k_\mu + \gamma^\mu \: i\sin k_\mu \right) }{(M_\chi-\sum_\nu \cos k_\nu)^2 + \sum_\nu (\sin k_\nu)^2} \ \left(-\sigma^y \right)
\end{eqnarray*}
\vspace{.3cm}
\begin{eqnarray*}
\hspace{1.8cm}
\parbox{39mm}{
\begin{fmffile}{zzz-V_M}
\begin{fmfgraph*}(25, 25)
\fmfleftn{l}{5}\fmfrightn{r}{3}
\fmf{plain,label.side=right,label=$J$}{l1,m}
\fmf{plain,label.side=right,label=$I$}{m,l5}
\fmf{photon,label.side=right,label=$q$}{r2,m}
\fmflabel{$k-q/2$}{l2}
\fmflabel{$k+q/2$}{l4}
\fmflabel{$\mu, IJ$}{r2}
\fmfdot{m}
\momentumarrow{a}{up}{4}{r2,m}
\fmffreeze
\fmf{phantom}{l1,m1}
\fmf{phantom,tension=3}{m1,m}
\fmf{phantom}{m2,l5}
\fmf{phantom,tension=3}{m,m2}
\momentumarrow{b}{left}{7}{l1,m1}
\momentumarrow{c}{left}{7}{m2,l5}
\end{fmfgraph*}
\end{fmffile}
}
= \ i\, e^{-iq_\mu/2} \left(-\sigma^y \right) \left(\gamma^\mu i\cos k_\mu + \sin k_\mu \right)
\end{eqnarray*}
\vspace{.3cm}
\begin{eqnarray*}
\hspace{1cm}
\parbox{36mm}{
\begin{fmffile}{zzz-U_M}
\begin{fmfgraph*}(20, 10)
\fmfleftn{l}{3}\fmfrightn{r}{3}
\fmf{photon,label.side=left,label=$q \ \mathrm{mod} \: 2\pi$}{l2,r2}
\fmfdot{l2,r2}
\fmflabel{$\mu, IJ \ $}{l2}
\fmflabel{$\ \nu, J'I'$}{r2}
\momentumarrow{a}{up}{4}{r2,l2}
\end{fmfgraph*}
\end{fmffile}
}
= \ \frac{T}{2} \: \delta_{\mu\nu} \left(\delta_{IJ'} \delta_{JI'} - \delta_{II'}\delta_{JJ'} \right)
\end{eqnarray*}
where the undirected lines are the $\psi^I$ Majorana fermions and the wavy lines are the real, antisymmetric $\theta_{IJ}$ (Higgsed) gauge fields. It is clear that the self-energy is just that of the Dirac case with $N_f$ replaced by $(N_f-1)/2$, so $-\Sigma(p=0) \approx -0.113 \, (N_f-1)T/2$.

We are left with the Majorana case with $N_f=N-1$, at which there is an additional unconfined Majorana fermion $\chi'$. The real $\eta_{N_f I}$ fields which connect the $\psi$ sector to the $\chi'$ sector must also be taken into account. The new Feynman rules are
\begin{eqnarray*}
\parbox{30mm}{
\begin{fmffile}{zzz-G_M1}
\begin{fmfgraph*}(25, 10)
\fmfleftn{l}{3}\fmfrightn{r}{3}
\fmf{dbl_plain,label.side=right,label=$k$}{r2,l2}
\momentumarrow{a}{up}{4}{r2,l2}
\fmfdot{l2,r2}
\end{fmfgraph*}
\end{fmffile}
}
= \ \frac{-M_\chi+\sum_\mu \left(\cos k_\mu + \gamma^\mu \: i\sin k_\mu \right) }{(M_\chi-\sum_\nu \cos k_\nu)^2 + \sum_\nu (\sin k_\nu)^2} \ \left(-\sigma^y \right)
\end{eqnarray*}
\vspace{.3cm}
\begin{eqnarray*}
\hspace{1.8cm}
\parbox{40mm}{
\begin{fmffile}{zzz-V_M1}
\begin{fmfgraph*}(25, 25)
\fmfleftn{l}{5}\fmfrightn{r}{3}
\fmf{plain,label.side=right,label=$I$}{l1,m}
\fmf{dbl_plain}{m,l5}
\fmf{dbl_wiggly,label.side=right,label=$q$}{r2,m}
\fmflabel{$k-q/2$}{l2}
\fmflabel{$k+q/2$}{l4}
\fmflabel{$\mu, N_f I$}{r2}
\fmfdot{m}
\momentumarrow{a}{up}{5}{r2,m}
\fmffreeze
\fmf{phantom}{l1,m1}
\fmf{phantom,tension=3}{m1,m}
\fmf{phantom}{m2,l5}
\fmf{phantom,tension=3}{m,m2}
\momentumarrow{b}{left}{7}{l1,m1}
\momentumarrow{c}{left}{7}{m2,l5}
\end{fmfgraph*}
\end{fmffile}
}
= \ i\, e^{-iq_\mu/2} \left(-\sigma^y \right) \left(\gamma^\mu i\cos k_\mu + \sin k_\mu \right)
\end{eqnarray*}
\vspace{.3cm}
\begin{eqnarray*}
\hspace{1cm}
\parbox{37mm}{
\begin{fmffile}{zzz-U_M1}
\begin{fmfgraph*}(20, 10)
\fmfleftn{l}{3}\fmfrightn{r}{3}
\fmf{dbl_wiggly,label.side=left,label=$q \ \mathrm{mod} \: 2\pi$}{l2,r2}
\fmfdot{l2,r2}
\fmflabel{$\mu, N_f I \ $}{l2}
\fmflabel{$\ \nu, I' N_f$}{r2}
\momentumarrow{a}{up}{5}{r2,l2}
\end{fmfgraph*}
\end{fmffile}
}
= \ T \: \delta_{\mu\nu} \: \delta_{II'}
\end{eqnarray*}
where the double lines are the $\chi'$ fermion and the double wavy lines are the real $\eta_{N_f I}$ (Higgsed) gauge fields. The self-energy of the $\psi^I$ fermions becomes $-\Sigma(p=0) \approx -0.113 \, (N_f+1)T/2 = -0.113 \, NT/2$, while that of the $\chi'$ fermion is $-\Sigma'(p=0) \approx -0.113 \, N_f = -0.113 \, (N-1)$.

\section{Lattice and/or Yang-Mills Regularization} \label{sec:YM} 

\subsection{Physical Argument for IR Equivalence} \label{sec:phaseTransitions}

In this paper, we regularized Chern-Simons-matter theories by realizing them on a lattice. There is, however, another regularization scheme that is commonly used in the continuum: the addition of a Yang-Mills (YM) term with a coupling constant $g$. In particular, the fixed points of interest may be defined as the IR fixed points of UV gauge theories with $g^2$ far below the cutoff scale, which can be regularized by imposing perturbative renormalization conditions. As one flows to the IR, one expects $g^2\to\infty$, as the YM term is irrelevant. However, it softens the UV behavior of the gauge field, as is clear from its propagator.

In contrast, on the lattice, there is no compelling reason for us to include the YM term (indeed, in the main text we took $g^2\rightarrow\infty$ on the lattice), as the lattice suffices as a regulator. From this perspective, YM is no different from other irrelevant terms. However, this leads to the question of whether the two different regularization schemes flow to the same IR. Universality suggests an affirmative answer, especially since the YM term is dominated in the IR by the Chern-Simons (CS) interaction.\footnote{When there is no CS term, the YM term becomes the leading term and is important. For instance, lattice proofs of bosonic particle-vortex duality \cite{peskin:particleVortex,dasgupta:particleVortex} require a Maxwell term to access the critical point.} Nevertheless, a phase transition is not out of the question, especially since we are considering $g^2\ll 1$ and $g^2 \gg 1$. The purpose of this appendix is to confirm that no phase transition occurs as we decrease $g^2$.

The following simple argument suggests the two regularizations flow to the same IR. First suppose one uses the continuum YM regularization. If one starts the RG flow at a scale $\Lambda$, then there is some intermediate scale $\mu' \sim g^2 \Lambda$ at which the effective coupling, $g^2_{\mu'}$, runs to order $1$ in units of $\mu'$. That is, at the intermediate scale $\mu'$, one has the boson coupled to a CS and YM action, with an order $1$ YM coupling. Now suppose one uses our lattice construction. One gets the same for free. Let the intermediate scale $\mu'$ be such that $\mu' \lesssim 3-M_\chi$. Then one can integrate out the $\chi'$ fermion, which generates not only the desired CS term, but also a YM term with $g_{\mu'}^2 \simeq 3-M_\chi$, which is again order $1$ in units of $\mu'$. Therefore, using either regularization scheme, there is some intermediate energy scale $\mu'$ at which one has a boson coupled to CS and an order $1$ YM, which then flows to the same IR at $\mu\ll \mu'$. Of course, under these RG flows, we also generate an infinite set of other order-1 interactions, so we can never say that the lattice theory has become a CS+YM theory, but it is undeniable that at the scale $\mu'$ the two theories obtained by flowing from $\Lambda$ appear awfully similar.

The argument above suggests that either the lattice or the YM regularization would yield the same IR, but one can also consider combining these two regularizations. That is, instead of taking $g^2\rightarrow \infty$ on the lattice, we can take
\begin{align}
\mbox{IR energy scale of interest} \ \ll \ g^2 \ \ll \ 1 \ \equiv \ \mbox{Inverse lattice scale} \ .
\label{YM_scale}
\end{align}
We emphasize again that there is no compelling reason to do so for regularization purposes. We simply wish to explore the relationship between the different regularization schemes, and in particular to demonstrate that introducing a YM term alters neither our method nor our conclusions. That is, we will repeat the arguments of the main text in order to find a fixed point with $T\ll 1$.

Before proceeding, we note that there is the theoretical possibility of another phase transition, due to the existence of another dimensionless parameter in the UV: $T-T_c^0$, where $T_c^0$ is the critical temperature of the ungauged non-linear sigma model (which is the appropriate model to consider because $g^2\ll 1$). We will fall short of being able to address the behavior of our theories at such order-1 values of $T$. So, conservatively one might say that this paper demonstrates equivalences between free fermion (and Ising) fixed points and the IR limits of Chern-Simons theories coupled to scalars, but that these fixed points might not have the right to be called `gauged Wilson-Fisher' fixed points, since one cannot necessarily arrive at them by coupling Wilson-Fisher fixed points to gauge theories and then flowing to the IR (while tuning appropriately to hit the fixed point). Generally, this perspective seems rather conservative, as one can imagine increasing $T$, and simultaneously increasing $3-M_\chi$, so that one always has a critical theory; we then require only that this procedure does not dramatically alter the critical behavior at some $T$, and in particular that a `gauged Wilson-Fisher' fixed point exists.\footnote{One can easily argue, as in the main text, for the existence of a phase transition, but as $T$ is increased to be order-1 there is no proof that it remains second order. Furthermore, even if it is second order there is no proof that the fixed point exists for sufficiently small $g^2, |T-T_c^0|$ so that it can be considered a `gauged Wilson-Fisher' fixed point. That is, a weakly gauged lattice non-linear sigma model at a temperature near $T_c^0$ has an order-1 energy scale above which the scalars are not near their Wilson-Fisher fixed point, and so in order to have a `gauged Wilson-Fisher' fixed point in the strictest sense we must have $|T-T_c^0| \lesssim g^2 \ll 1$.\label{ft:gWF}} In the cases where we have found multiple fixed points, one may or may not be willing to make the analogous assumption that there are multiple corresponding such `gauged Wilson-Fisher' fixed points. Our arguments do not shed much light on this question; even if there is no phase transition, it is possible that one (or both) of our fixed points cannot satisfy the stringent conditions laid out in footnote \ref{ft:gWF}. What we hope to have clearly demonstrated is that it is overly pessimistic to use the usual Higgsing-down reasoning (which violates the conservative requirements of footnote \ref{ft:gWF}) in order to rule out the usual dualities in these cases; indeed, this Higgsing argument forces us to introduce large temperature deformations, and when one does so new degrees of freedom may become light.

Similarly, one might feel more comfortable studying the generation of CS terms by fermions with $3-M_\chi \gg g^2$. Unfortunately, for small enough $N$ our search for a fixed point (with $T,\, 3-M_\chi \ll 1$) fails if we demand this, so we do not assume it below. (This would presumably be rectified by including higher orders of $T,\, 3-M_\chi$ in perturbation theory.) For example, our solution for $T_c$ in the $U(N)$ case requires $3-M_\chi < 0.6 \, g^2N$, as can be seen by choosing some $M_\chi$ slightly below 3 and studying the limit $g^2/T\to 0$ in \eqref{eq:answer} (and adding to it the analogous contribution with $N_f\to N-N_f$ discussed in the preceding paragraph). We are nevertheless confident that our fermion implements the desired CS interaction, thanks to the parity anomaly.

Having said this, we now assume the condition \eqref{YM_scale} involving $g^2$, as well as the usual analogous assumption \eqref{arrange_scales} for $3-M_\chi$. We make no additional assumptions about the relationships between $g^2,\, T$, and $3-M_\chi$.

\subsection{Including Yang-Mills on the Lattice} \label{sec:YMcomputation}

The YM theory on the lattice is given by $e^{-S_{YM}}$ where (for a $U(N)$ theory)
\begin{align}
-S_{YM} =& \ \frac{1}{4g^2} \sum_{n, \mu, \nu} \tra\left( U_{n\nu}^\dagger U_{(n+\hat\nu)\mu}^\dagger U_{(n+\hat{\mu})\nu} U_{n\mu} - 1 \right).\label{eq:compact}
\end{align}
For small $g^2$, we can expand the $U_{ab}$ matrix using the Lie algebra elements $b_{ab}$:
\begin{align}
-S_{YM} = - \frac{1}{4g^2} \sum_{n, \mu, \nu} \left( \tra \left( b_{(n+\hat\mu)\nu} - b_{n\nu} - b_{(n+\hat\nu)\mu} + b_{n\mu} \right)^2 \ + \ \cdots \right) \ , \label{eq:noncompact}
\end{align}
where $(\ldots)$ are higher order terms.
\footnote{While this expansion seems innocent, we recall (see also footnote \ref{ft:noMaxwell}) that it actually dramatically changes the IR physics, since the new action only has the trivial $U=1$ saddle, whereas the original action had many saddles \cite{polyakov:compactLattice}. These saddles are characterized by the presence of Dirac strings (i.e., $2\pi$ flux tubes as narrow as one plaquette) which end on monopoles. Such saddles do not exist in continuum $U(1)$ gauge theories on $\mathbb{R}^3$, but they can exist on a lattice because the core of a monopole is a lattice cube, which is non-singular. 
This distinction is sometimes emphasized by calling the central $U(1)\subset U(N)$ `compact' when the action is \eqref{eq:compact} and `non-compact' when \eqref{eq:noncompact} is employed. For either a `compact' or `non-compact' $U(1)$ gauge field, a Dirac string is invisible to charged particles. However, in the former case, a Dirac string is also invisible in the Maxwell term, while in the latter it costs extra action in the Maxwell term, just like a thin solenoid. So the difference is really in the prescription of the Maxwell term.

The IR field theory dualities of interest require `non-compact' gauge fields, so we would really prefer to use \eqref{eq:noncompact}. We therefore add to \eqref{eq:compact} the following term that eliminates the extra monopole saddles:
\begin{align}
\frac{1}{4{g'}^2} \sum_{n, \mu, \nu} \frac{1}{2} \left( \arg \det \left( U_{n\nu}^\dagger U_{(n+\hat\nu)\mu}^\dagger U_{(n+\hat{\mu})\nu} U_{n\mu} \right) + 2\pi m_{n\mu\nu} \right)^2 \ . \label{eq:extra}
\end{align}
Here, $m_{n\mu\nu}=-m_{n\nu\mu}$ is a closed integer field on the plaquettes to be summed over in the path integral. `Closed' means the sum of the $m$ field coming out of the faces of each lattice cube must vanish; locally we may write the closed $m$ field as the lattice curl of some integer gauge field $m'$ on the link, and we can combine the central part of $b$ with $2\pi m'$ into a real gauge field. This explains the historical name `non-compact', but note the real gauge field only makes sense locally. The reason we demand $m$ to be closed is the following: if we relax the closedness condition of $m$, then the action will be the Villainized version of the `compact' Maxwell term \eqref{eq:compact} for the central part of the gauge field. In this Villainized `compact' Maxwell, the integer $m$ field is interpreted as the Dirac strings, whose end points (i.e. lattice cubes out of which $m$ is not closed) are fluctuating monopoles, so that the total flux (the inside of the parenthesis of \eqref{eq:extra}) appears non-conserved. To keep the computations below unaltered, we can take $1/{g'}^2 \ll 1/g^2$ and ignore \eqref{eq:extra} in the perturbative expansion. \label{ft:compact_non-compact}}
In our problem, suppose we have $N_f$ bosons at temperature $T$. Then the gauge field components $b_{aI}=b_{Ia}^\ast$ ($I=1,\ldots,N_f$) are Higgsed; the $IJ$ components are what we called the Hermitian matrix $\theta_{IJ}$ and the $AJ$ ($A = N_f+1, \ldots, N$) components are what we called the complex matrix $\eta_{AJ}$. Their perturbative action has the leading terms
\begin{align}
& - \frac{1}{4g^2} \sum_{n, \mu, \nu}  \left( \sum_{I, J} \left| \theta_{(n+\hat\mu)\nu} - \theta_{n\nu} - \theta_{(n+\hat\nu)\mu} + \theta_{n\mu} \right|_{IJ}^2 \ + \ \cdots \right) - \frac{1}{2T} \sum_{n, \mu} \left( \sum_{I,J} |\theta_{n\mu}|_{IJ}^2 + \dots \right) \nonumber \\[.2cm]
& - \frac{1}{4g^2} \sum_{n, \mu, \nu}  \left( 2\sum_{A, J} \left| \eta_{(n+\hat\mu)\nu} - \eta_{n\nu} - \eta_{(n+\hat\nu)\mu} + \eta_{n\mu} \right|_{AJ}^2 \ + \ \cdots \right) - \frac{1}{2T} \sum_{n, \mu} \left( \sum_{A, J} |\eta_{n\mu}|_{AJ}^2 + \dots \right).
\end{align}
On the other hand, there is a residual $U(N-N_f)$ gauge group with gauge field $b_{AB}$. This is coupled to the ${\chi'}^A$ fermions. Since the YM term for $b_{AB}$ is irrelevant, we expect this gauge field to confine $\chi'^A$ into massive bosons, as in the main text. That is, currents involving $\chi'$ vanish. In particular, these fermions should not contribute a CS term for the background field, $A$, in the IR.

Now we compute the self-energy of the $\psi^I$ fermion and show it is again positive and increases with $T$. The main change compared to Appendix \ref{sec:diagram} is the gauge field propagator:
\begin{eqnarray*}
&& \hspace{1.5cm}
\parbox{36mm}{
\begin{fmffile}{zzz-U_YM}
\begin{fmfgraph*}(20, 10)
\fmfleftn{l}{3}\fmfrightn{r}{3}
\fmf{photon,label.side=left,label=$q \ \mathrm{mod} \: 2\pi$}{l2,r2}
\fmfdot{l2,r2}
\fmflabel{$\mu, IJ \ $}{l2}
\fmflabel{$\ \nu, J'I'$}{r2}
\momentumarrow{a}{up}{4}{r2,l2}
\end{fmfgraph*}
\end{fmffile}
} \\[.3cm]
&= & \ \delta_{II'}\delta_{JJ'} \left[ \frac{\delta_{\mu\nu}}{T} + \frac{\delta_{\mu\nu}\, \sum_\lambda \left|e^{iq_\lambda}-1\right|^2 - \left(e^{iq_\mu}-1\right)\left(e^{-iq_\nu}-1\right)}{g^2} \right]^{-1} \\[.3cm]
&= & \ \delta_{II'}\delta_{JJ'} \left[ \frac{g^2}{g^2/T + \sum_\lambda \left|e^{iq_\lambda}-1\right|^2} \left( \delta_{\mu\nu} - \frac{\left(e^{iq_\mu}-1\right)\left(e^{-iq_\nu}-1\right)}{\sum_\lambda \left|e^{iq_\lambda}-1\right|^2} \right) \right. \nonumber \\[.2cm]
&& \ \left. \hspace{1.8cm} + T\: \frac{\left(e^{iq_\mu}-1\right)\left(e^{-iq_\nu}-1\right)}{\sum_\lambda \left|e^{iq_\lambda}-1\right|^2} \right] \ .
\end{eqnarray*}
Note the following features of this propagator. First, it does not diverge in the IR, due to the Higgs mechanism; this is related to the locality of the theory that obtains after integrating out the gauge field. Second, the fact that the second term does not vanish as $g^2\to 0$ is responsible for the fact that we cannot perturbatively compute the fermion self-energy correction when $T$ is of order 1. Choosing a gauge besides unitary gauge can ameliorate this problem, but then one must compute the correlation function $\avg{\phi^\dagger_n \chi_n \phi_{n'} \bar\chi_{n'}}$ in a theory where $\phi$ is dynamical and $T$ is order 1.

To leading order in $(T, g^2)$, the self-energy diagrams for the $\psi^I$ fermions are
\begin{eqnarray*}
\parbox{32mm}{
\begin{fmffile}{zzz-YMSigma0}
\begin{fmfgraph*}(33, 20)
\fmfleftn{l}{4}\fmfrightn{r}{4}
\fmf{fermion,label.side=left}{m,l2}
\fmf{fermion,label.side=left}{r2,m}
\fmffreeze
\fmf{phantom,tension=5}{m1,l4}
\fmf{phantom,tension=5}{m1,r4}
\fmf{phantom}{m1,m}
\fmf{photon,left=.9}{m,m1}
\fmf{photon,left=.9}{m1,m}
\fmfdot{m}
\end{fmfgraph*}
\end{fmffile}
}
\hspace{.9cm}
\parbox{32mm}{
\begin{fmffile}{zzz-YMSigma2}
\begin{fmfgraph*}(33, 20)
\fmfleftn{l}{4}\fmfrightn{r}{4}
\fmf{fermion,label.side=left}{m1,l2}
\fmf{fermion,tension=0.5}{m2,m1}
\fmf{fermion,label.side=left}{r2,m2}
\fmffreeze
\fmf{photon,left=1}{m1,m2}
\fmfdot{m1,m2}
\end{fmfgraph*}
\end{fmffile}
}
\hspace{.9cm}
\parbox{32mm}{
\begin{fmffile}{zzz-YMSigma00}
\begin{fmfgraph*}(33, 20)
\fmfleftn{l}{4}\fmfrightn{r}{4}
\fmf{fermion,label.side=left}{m,l2}
\fmf{fermion,label.side=left}{r2,m}
\fmffreeze
\fmf{phantom,tension=5}{m1,l4}
\fmf{phantom,tension=5}{m1,r4}
\fmf{phantom}{m1,m}
\fmf{dbl_wiggly,left=.9}{m,m1}
\fmf{dbl_wiggly,left=.9}{m1,m}
\fmfdot{m}
\end{fmfgraph*}
\end{fmffile}
}
\hspace{.9cm}
\parbox{32mm}{
\begin{fmffile}{zzz-YMSigma22}
\begin{fmfgraph*}(33, 20)
\fmfleftn{l}{4}\fmfrightn{r}{4}
\fmf{fermion,label.side=left}{m1,l2}
\fmf{dbl_plain,tension=0.5}{m2,m1}
\fmf{fermion,label.side=left}{r2,m2}
\fmffreeze
\fmf{phantom_arrow,tension=0}{m2,m1}
\fmf{dbl_wiggly,left=1}{m1,m2}
\fmfdot{m1,m2}
\end{fmfgraph*}
\end{fmffile}
} \ .
\end{eqnarray*}
We have to include the $\chi'$ fields (arrowed double lines) in the internal lines, as in contrast to the original $g^2\rightarrow \infty$ case where they confine on each lattice link, now, above the scale $g^2$, the $\chi'$ fermions exist as propagating fermions. The propagator of the $\chi'$ fermion is the same as that of the $\psi$ fermion. The $\eta_{AI}$ propagator (double wavy lines) is the same as the $\theta_{IJ}$ propagator expect for $\delta_{II'}\delta_{JJ'} \rightarrow \delta_{II'}\delta_{AA'}$ and $T\rightarrow 2T$. In Appendix \ref{sec:diagram} the first and third diagrams which renormalize $M_\chi$ to $M_\psi$ were already accounted for in the real space exact mapping; now we have to compute them in momentum space. The $\bar\psi\psi \theta\theta$ vertex is given by
\begin{eqnarray*}
\hspace{2.7cm}
\parbox{37mm}{
\begin{fmffile}{zzz-V_YM2}
\begin{fmfgraph*}(25, 25)
\fmfleftn{l}{5}\fmfrightn{r}{5}
\fmf{fermion,label.side=right,label=$J$}{l1,m}
\fmf{fermion,label.side=right,label=$I$}{m,l5}
\fmf{photon,tension=0.35,label.side=right,label=$q$}{m,r2}
\fmf{photon,tension=0.35,label.side=right,label=$q'$}{r4,m}
\fmflabel{$k-(q-q')/2 \!\!\!\!\!$}{l2}
\fmflabel{$k+(q-q')/2 \!\!\!\!\!$}{l4}
\fmflabel{$\nu, K'J'$}{r2}
\fmflabel{$\mu, I'K$}{r4}
\fmfdot{m}
\fmffreeze
\momentumarrow{b}{down}{4}{r2,m}
\momentumarrow{c}{up}{4}{m,r4}
\end{fmfgraph*}
\end{fmffile}
}
= \ \frac{\delta^{\mu\nu}}{2} \: \delta_{II'}\, \delta_{JJ'}\, \delta_{KK'} \, e^{-i(q-q')_\mu/2} \left(\gamma^\mu i\sin k_\mu - \cos k_\mu \right)
\end{eqnarray*}
(which is $\partial_{k_\mu} \partial_{k_\nu}/2$ of the inverse propagator, by gauge invariance) and the $\bar\psi\psi \eta^\dagger \eta$ vertex is obtained by replacing $\delta_{KK'}$ with $\delta_{AA'}$. In YM theory there are also interactions among the gauge fields, but thanks to the vanishing of any tadpole diagram, we do not need to include them to first order in $(T, g^2)$.

Let's compute the first two diagrams with external $p=0$ (the last two diagrams are computed the same way with $T\rightarrow 2T$ and $N_f \rightarrow N-N_f$). Gauge invariance leads to Ward identities which imply that the $\left(e^{iq_\mu}-1\right)\left(e^{-iq_\nu}-1\right)$ terms in the gauge field propagator make $\mathcal{O}(g^2(3-M_\chi)) + \mathcal{O}(T(3-M_\chi))$ contributions at $p=0$ upon summation of the diagrams,\footnote{The Ward identities are the following. Let $G(k)$ be the propagator,  $\Gamma^\mu(k; q)$ be the single gauge field vertex, and $\Xi^{\mu\nu}(k; q, q')$ be the double gauge field vertex. Then $\sum_\mu (e^{iq_\mu}-1) \Gamma^\mu(k; q) = G^{-1}(k+q/2) - G^{-1}(k-q/2)$ and $2\sum_\nu (e^{iq_\nu}-1) \Xi^{\mu\nu}(k; q', q) =  \Gamma^\mu(k+q/2; q') - \Gamma^\mu(k-q/2; q')$.} so we only need the $\delta^{\mu\nu}$ term of the gauge field propagator. The first two diagrams at $p=0$ contribute
\begin{align}
& \delta_{IJ} \, N_f \, \int_{-\pi}^\pi \frac{d^3 k}{(2\pi)^3} \ \frac{g^2}{g^2/T + 2\sum_\lambda \left( 1-\cos k_\lambda \right)} \ \sum_\mu \left[ -\frac{1}{2} + \frac{\cos k_\mu \: (M_\chi-\sum_\lambda \cos k_\lambda) - (\sin k_\mu)^2}{(M_\chi-\sum_\kappa \cos k_\kappa)^2 + \sum_\kappa (\sin k_\kappa)^2} \right] \nonumber \\
= & \  \delta_{IJ} \, N_f \, \int_{-\pi}^\pi \frac{d^3 k}{(2\pi)^3} \ \frac{g^2}{g^2/T + 2\sum_\lambda \left( 1-\cos k_\lambda \right)} \ \brackets{ -\frac{5}{2} + \frac{M_\chi ( M_\chi - \sum_\lambda \cos k_\lambda)}{(M_\chi - \sum_\kappa \cos k_\kappa)^2 + \sum_\kappa (\sin k_\kappa)^2 } } \ , \label{eq:answer}
\end{align}
where in the square bracket in the first line, the $-1/2$ is from the first diagram and the rest, from the second diagram, is the same as \eqref{self-energy_calculation}. The integrand is always negative, and moreover, its $T$ derivative is also negative. Since the diagrams compute $-\Sigma$, this means the total self-energy, at leading order in $(3-M_\chi, T, g^2)$, is always positive and increases with $T$ (towards $\sim 0.6\, g^2 N$). This is the same as in the $g^2\rightarrow \infty$ case of the main text. There is no substantial change if we consider the $SO(N)$ theories instead of $U(N)$. Therefore, our results in this paper are independent of whether or not we have a weakly coupled YM term in the UV, in agreement with our intuitive argument given above.

\bibliography{nAb_duality}{}
\bibliographystyle{utphys}
\end{document}